\documentclass[twocolumn]{revtex4}
\usepackage[colorlinks,citecolor=blue]{hyperref}
\usepackage{textcomp}
\usepackage{subfigure}
\usepackage{verbatim}
\usepackage{bm}
\usepackage{hyperref}
\usepackage{amsmath}
\usepackage{graphicx}
\usepackage{epstopdf}
\usepackage{amssymb}
\usepackage{extarrows}
\usepackage{color}
\usepackage{CJK}
\usepackage{cancel}
\usepackage{ulem}
\usepackage[utf8]{inputenc}
\usepackage{url}
\usepackage{lineno}
\newcommand{\ba}{\begin{eqnarray}}
\newcommand{\ea}{\end{eqnarray}}
\newcommand{\be}{\begin{equation}}
\newcommand{\ee}{\end{equation}}
\newcommand{\gr}{\mathrm{GR}}
\newcommand{\gw}{\mathrm{GW}}

\newcommand{\m}{\mathrm{max}}

\newcommand{\oct}{\mathrm{oct}}

\newcommand{\au}{\mathrm{AU}}

\newcommand{\IN}{\mathrm{in}}

\def\e1{e_1^2}

\definecolor{ochre}{rgb}{0.8, 0.47, 0.13}

\begin{document}
\title{Uncovering a hidden black hole binary from secular eccentricity variations \\of a tertiary star}
\author{Bin Liu}
\email{bin.liu@nbi.ku.dk}
\affiliation{Niels Bohr International Academy, Niels Bohr Institute, Blegdamsvej 17, 2100 Copenhagen, Denmark}
\author{Daniel J. D'Orazio}
\affiliation{Niels Bohr International Academy, Niels Bohr Institute, Blegdamsvej 17, 2100 Copenhagen, Denmark}
\author{Alejandro Vigna-G\'omez}
\affiliation{Niels Bohr International Academy, Niels Bohr Institute, Blegdamsvej 17, 2100 Copenhagen, Denmark}
\author{Johan Samsing}
\affiliation{Niels Bohr International Academy, Niels Bohr Institute, Blegdamsvej 17, 2100 Copenhagen, Denmark}

\begin{abstract}
{We study the dynamics of a solar-type star orbiting around a black hole binary (BHB) in a nearly coplanar system.
We present a novel effect that can prompt a growth and significant oscillations of the eccentricity of the stellar orbit
when the system encounters an ``apsidal precession resonance", where the apsidal
precession rate of the outer stellar orbit matches that of the inner BHB.
The eccentricity excitation requires the inner binary to have a  non-zero eccentricity and unequal masses,
and can be created even in non-coplanar triples.
We show that the secular variability of the stellar orbit's apocenter, induced by the changing eccentricity, could be potentially detectable by \textit{Gaia}. Detection is favorable for BHBs emitting gravitational waves in the frequency band of the Laser Interferometer Space Antenna (LISA), hence providing a distinctive, multi-messenger probe on the existence of stellar-mass BHBs in the Milky Way. 
}
\end{abstract}

\maketitle

\section{Introduction}

About 90 double compact object merger events have been detected by the LIGO-Virgo-KAGRA Collaboration in the first three observing runs \cite{LIGO-2021}.
The avenues to produce stellar-mass
black-hole binary (BHB) mergers include different formation channels and environments.
Some of them are
isolated binary evolution
\cite{Lipunov-1997,Lipunov-2007,Podsiadlowski-2003,Belczynski-2010,Belczynski-2016,Dominik-2012,Dominik-2013,Dominik-2015,Alejandro-2017},
chemically homogeneous evolution \cite{Mandel-2016,Marchant-2016,duBuisson,Riley}, and multiple-body evolution in the gas disks of active galactic nuclei
\cite{Baruteau-2011,McKernan-2012,McKernan-2018,Bartos-2017,Stone-2017,Leigh-2018,Secunda-2019,Yang-2019,Grobner-2020,Ishibashi-2020,Tagawa-2020,
Liyaping-2021,Ford-2021,Samsing-Nature,Lirixin-2022,Lijiaru-2022}.
Additionally,there are various flavors of dynamical channels that involve either strong gravitational scatterings in dense clusters
\cite{Zwart(2000),OLeary(2006),Miller(2009),Banerjee(2010),Downing(2010),Ziosi(2014),Rodriguez(2015),Samsing(2017),
Samsing(2018),Rodriguez(2018),Gondan(2018)}, tertiary-induced mergers
via von Zeipel-Lidov-Kozai (ZLK) oscillations
\cite{vonZeipel,Lidov,Kozai,Smadar,Miller-2002,Wen-2003,Antonini-2012,Antonini(2017),Silsbee(2017),Petrovich-2017,Liu-ApJ,Xianyu-2018,
Hoang-2018,Liu-Quadruple,Fragione-Quadruple,Fragione-nulearcluster,Zevin-2019,Liu-HierarchicalMerger}
or flyby-induced mergers \cite{Michaely-2019,Michaely-2020}.
However, the relative contribution of each channel and the astrophysical origin of the detected mergers is still unclear.

BHB progenitors are expected to be numerous but remain undetected as an abundant population in our Universe.
Searching for these inspiraling BHBs is of great importance to understand the origin of gravitational-wave (GW) sources.
If the BHBs are not accreting, these quiescent sources can be detected via GWs.
Since BHBs in the inspiral phase
are still far from merger,
the associated GWs are in the low-frequency band that can be explored by future spaceborne GW observatories,
such as LISA \cite{LISA}, TianQin \cite{TianQin}, Taiji \cite{TaiJi}, B-DECIGO \cite{DECIGO},
Decihertz Observatories \cite{DeciHZ}, and TianGO \cite{TianGo}.
Also, since a significant fraction of compact BHBs may be members of hierarchical systems \cite{Tokovinin,Raghavan,Fuhrmann},
the motion of a nearby visible object (such as a star or a pulsar)
can be used to search for BHBs.
In this scenario, the inner BHB can perturb the outer orbit,
either inducing short-term orbital oscillations (tertiary orbit becomes quasi-Keplerian),
or causing long-term oscillations of the eccentricity and the orientation of the angular momentum
when the tertiary orbit is highly inclined \cite{Naoz-2017,Chiang-2018,Naoz-2019}.
If the tertiary object is bright enough, radial velocity measurements
could be used to determine the short and long term deviations from a Keplerian orbit \cite{Suto-1,Suto-2,Suto-3}.

Currently, observations show that most triple-star systems are less inclined or nearly coplanar \cite{Ransom,Thompson,Eisner}.
Here, we consider a solar-type star orbiting around a BHB
and study the secular evolution
of the stellar orbit when the triple system is coplanar.
We show that the outer binary may experience an eccentricity growth
driven by the ``apsidal precession resonance" \cite{Liu-Yuan,Liu-2020-PRD}.
Compared to the complex evolution of the orientation of the angular momentum (i.e., precession or nutation),
the secular change of the eccentricity of the outer stellar orbit
could provide distinctive evidence to reveal the presence of a BHB.

\section{Apsidal precession resonance}

We consider an inner BHB with masses $m_1$, $m_2$,
and a solar-type star ($m_\star=1\ \rm{M_\odot}$) that moves around the center of mass of the inner bodies.
The reduced mass for the inner binary is $\mu_\IN\equiv m_1m_2/m_{12}$, with $m_{12}\equiv m_1+m_2$.
Similarly, the outer binary has $\mu_\star\equiv(m_{12}m_\star)/(m_{12}+m_\star)$.
The semi-major axes and eccentricities are denoted by $a_\IN$, $a_\star$ and $e_\IN$, $e_\star$, respectively.
The orbital angular momenta of two orbits are thus given by
$\textbf{L}_\IN=\mathrm{L}_\IN\hat{\textbf{L}}_\IN=\mu_\IN\sqrt{G m_{12}a_\IN(1-e_\IN^2)}\,\hat{\textbf{L}}_\IN$
and $\textbf{L}_\star=\mathrm{L}_\star\hat{\textbf{L}}_\star=\mu_\star\sqrt{G (m_{12}+m_\star)a_\star(1-e_\star^2)}\,\hat{\textbf{L}}_\star$.

When the triple system is less inclined or nearly coplanar,
the ZLK oscillations are not allowed to occur, but a significant eccentricity excitation of the inner binary may still be induced
\cite{Ford,Naoz-PN}.
A secular, ``apsidal precession resonance"
plays a dominant role if the total apsidal precession
of the inner binary matches the precession rate of the outer binary \cite{Liu-Yuan,Liu-2020-PRD}.
Precession of both inner and outer binaries
is driven by Newtonian and general relativistic (GR) effects .
Such resonance allows efficient angular momentum exchange between the inner and outer binaries.

\begin{figure*}
\begin{centering}
\includegraphics[width=6.4cm]{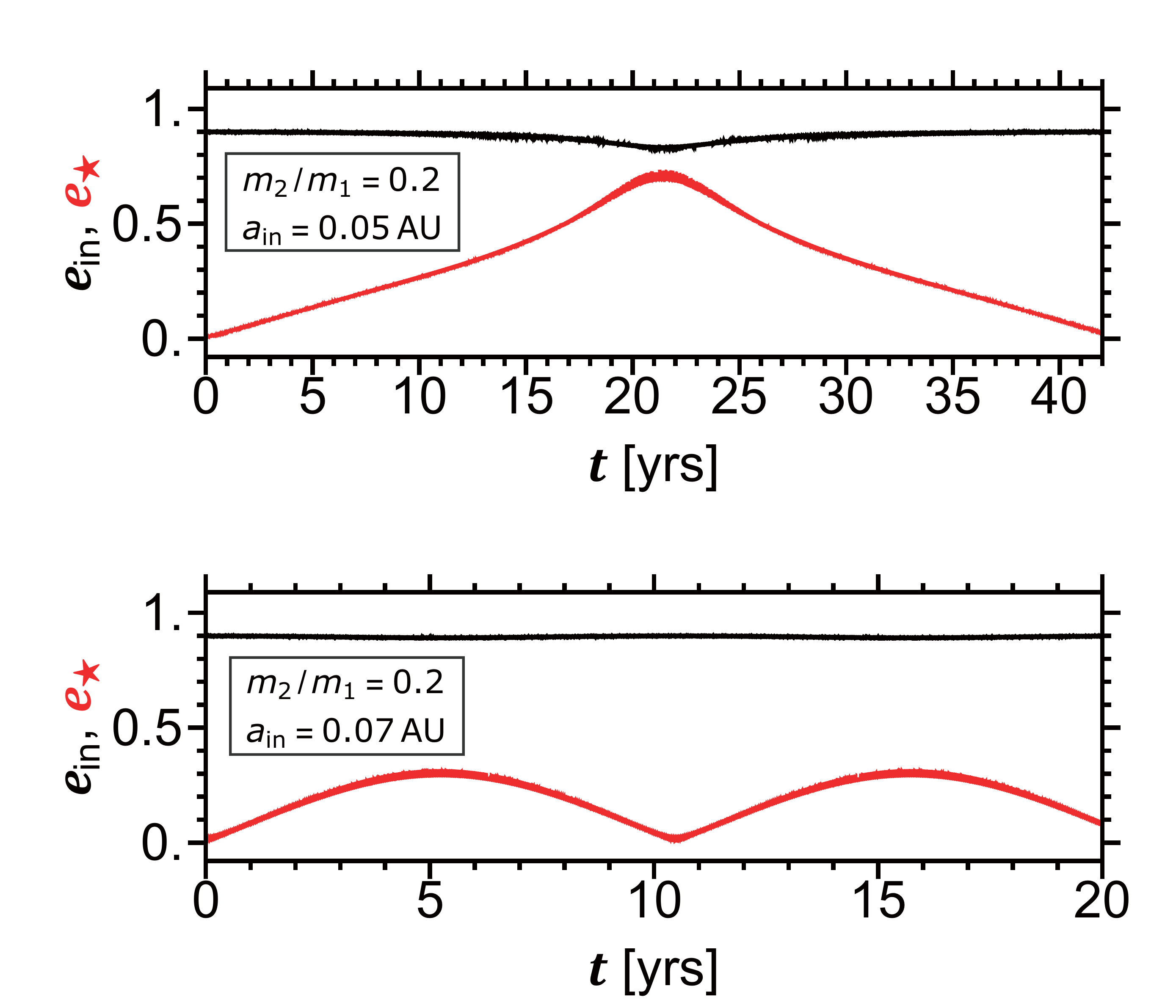}
\includegraphics[width=5.6cm]{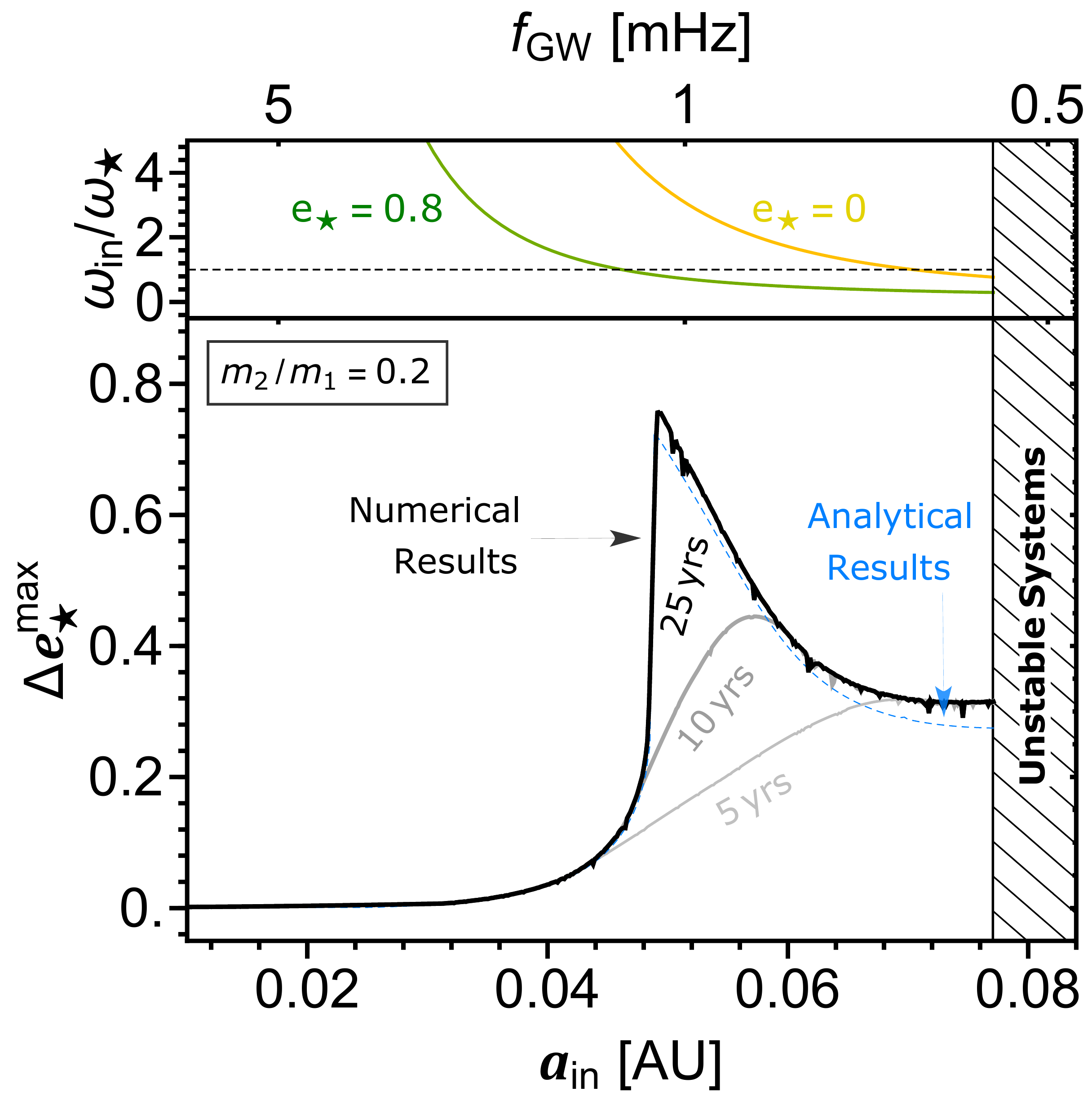}
\includegraphics[width=5.6cm]{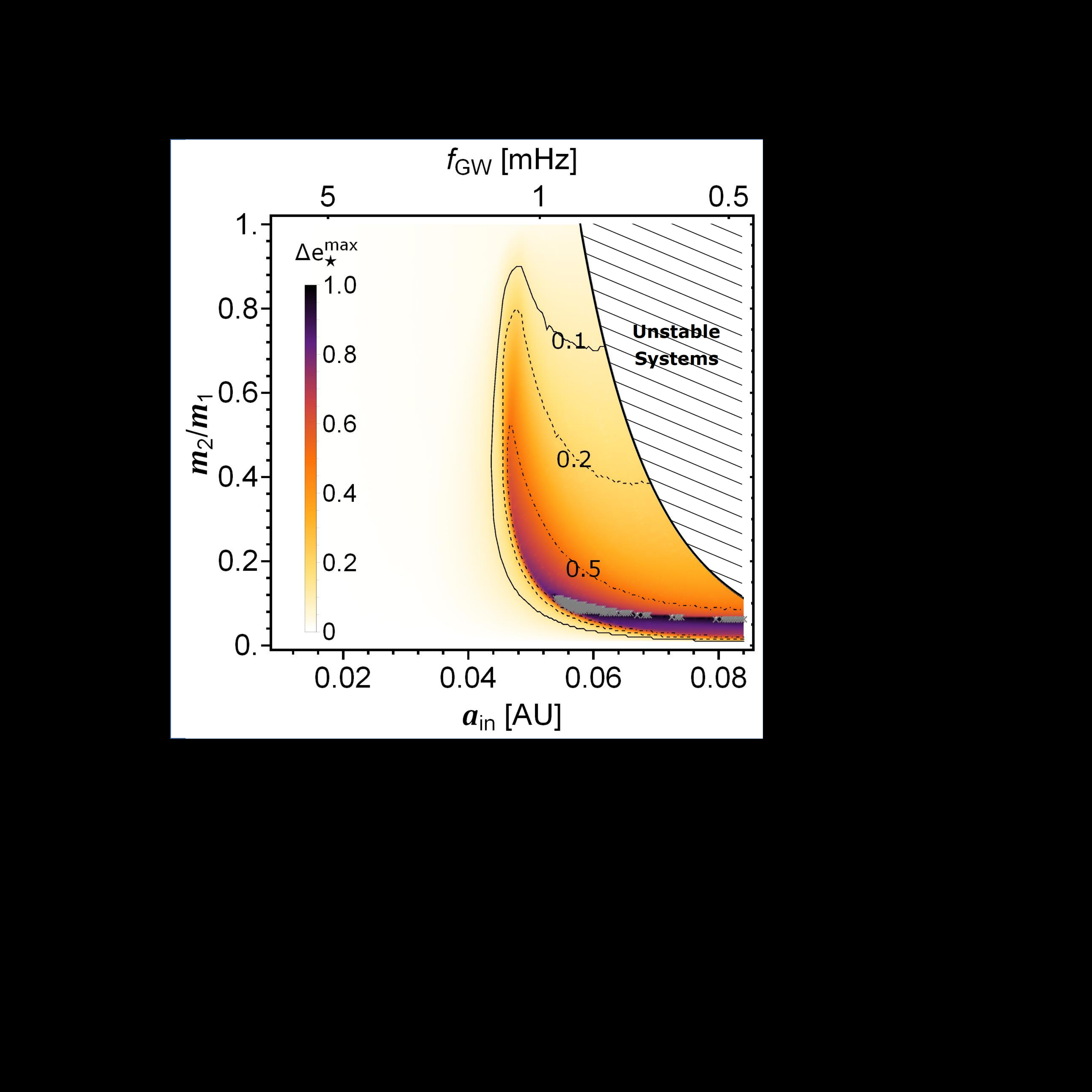}
\caption{Apsidal precession resonance in coplanar triple systems where the outer orbital period ($P_\star$) is set to be 15 days.
All the results are obtained by integrating the SA secular equations including GR effects (but without GW emission).
{\em Left panel:} Evolution examples of the orbital eccentricities in the inner (black) and outer (red) binaries, with $m_{12}=50\rm{M}_\odot$ and initial
$e_\IN^0=0.9$ and $e_\star^0=0$.
{\em Middle panel:} The ratio of apsidal precession rates (Equations \ref{eq:A11}-\ref{eq:A22};
including the dependence of finite eccentricities, i.e., $e_\IN^0$, $e_\star>0$) and
the maximum change of eccentricity $\Delta e_\star^\m$ of the outer stellar orbit as a function of the semimajor axis $a_\IN$.
The cross-hatched region corresponds to dynamically unstable triple systems \cite{Kiseleva-1996}.
The parameters are the same as in the left panel, except $a_\IN$ is relaxed to a range of values.
The labeled GW frequency is
the peak frequency at pericenter \cite{Wen-2003}.
The numerical results are obtained by integrating the SA secular equations over several timescales (as labeled), and
the analytical result is given by the energy and angular momentum conservation laws (dashed blue line).
{\em Right panel:} $\Delta e_\star^\m$ induced by the apsidal precession resonance in the ($m_2/m_1-a_\IN$) plane.
We use the same parameters as in the middle panel, taking into account the full range of mass ratios of the BHB.
The three black contours (solid, dashed and dot-dashed) specify $\Delta e_\star^\m=0.1, 0.2$ and $0.5$, respectively.
The gray crosses indicate the significant change of $e_\star$ that leads to instability.
}
\label{fig:Evolution}
\end{centering}
\end{figure*}

Here, we extend our previous studies to an ``inverse" secular problem, 
and address the question of how the apsidal precession resonance modifies the eccentricity evolution of the tertiary for the first time.
Since we are interested in the long-term orbital evolution, we adopt the single-averaged (SA; only averaging over
the inner orbital period) secular equations of motion,
taking into account the contributions from the Newtonian effect up to the octupole level of approximation
and the leading order GR effect in both inner and outer orbits.
The explicit SA equations are provided in \cite{Liu-ApJ,Blanchet}.

In Figure \ref{fig:Evolution}, the left panels show the evolution of eccentricities of both inner and outer
orbits. Starting with the eccentric inner binary, we see that $e_\star$
can be excited from the circular orbit and undergoes oscillations.
In the middle panel, we find that a large fraction of systems can develop eccentricities
($0.04\lesssim a_\IN/\au\lesssim0.08$), and an evident peak, $\Delta e_\star^\m\simeq0.8$,
can be resolved when the evolution is sufficiently long ($\gtrsim10$yrs).
The right panel illustrates the level of $e_\star-$excitation in the ($m_2/m_1-a_\IN$) plane.
The eccentricity of the stellar orbits can be excited for $a_\IN\gtrsim0.04\ \au$, and the
systems with smaller mass ratio tend to have larger $\Delta e_\star^\m$.
This is because the evolution of $e_\star$ is determined by the octupole-order secular interactions,
which can be quantified by terms proportional to \cite{Smadar}:
\be\label{eq:Oct}
\varepsilon_\oct\equiv\frac{m_1-m_2}{m_{12}}\frac{a_\IN}{a_\star}\frac{e_\star}{1-e_\star^2}.
\ee
We see that the eccentricities of some outer orbits reach significantly large $\Delta e_\star^\m$ close to unity, leading to unbound orbits.

\begin{figure*}
\begin{centering}
\includegraphics[width=5.5cm]{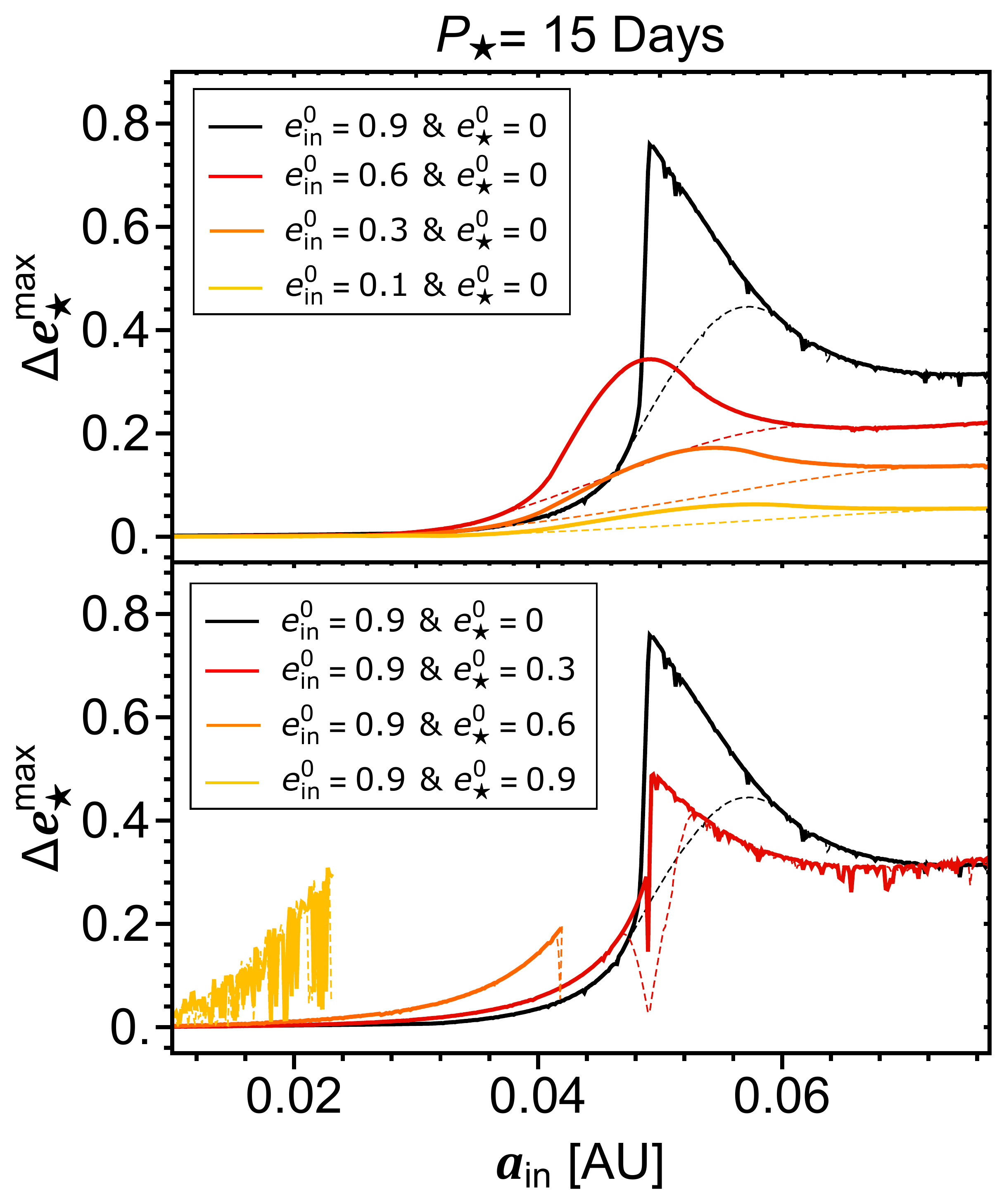}
\includegraphics[width=5.5cm]{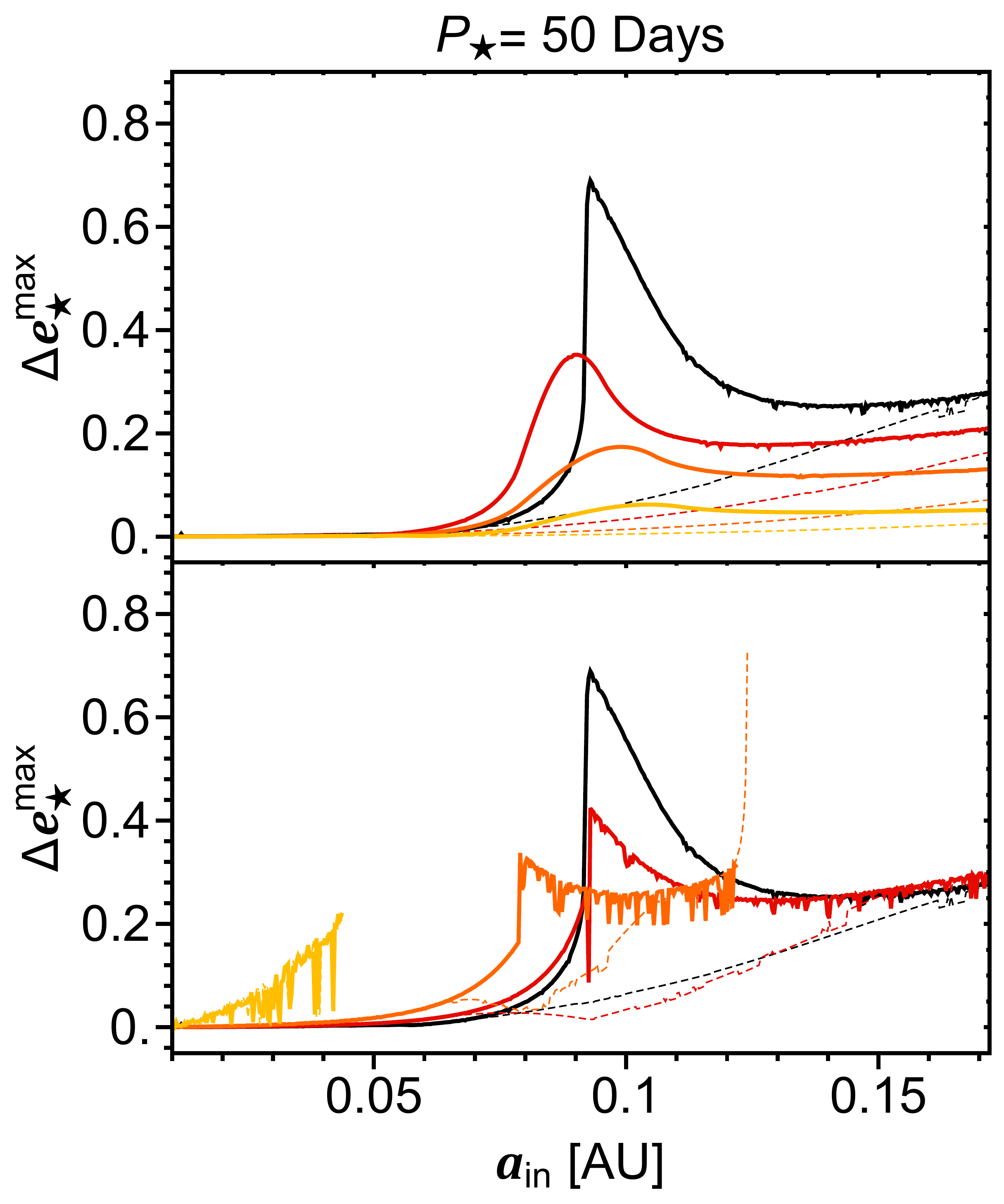}
\includegraphics[width=5.5cm]{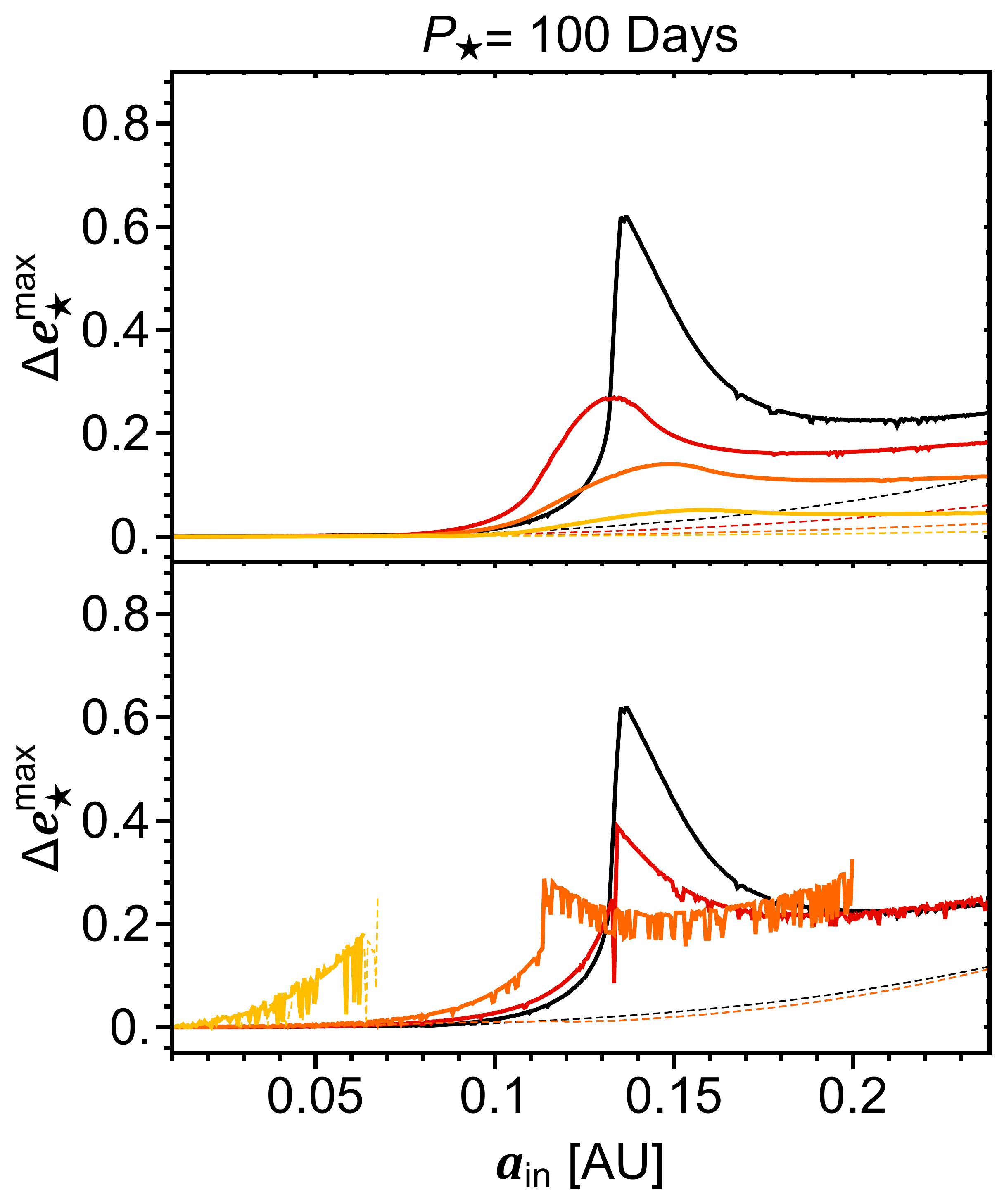}
\caption{
The maximum change of eccentricity as a function of the inner binary
semimajor axis.
We consider the inner BHB with the total mass $m_{12}=50\ \rm{M}_\odot$ and mass ratio $m_2/m_1=0.2$
(same as the middle panel of Figure \ref{fig:Evolution}),
and the stellar orbits with different orbital period (as labeled).
The results (solid lines) are obtained by the numerical integrations of SA equations
for 25 yrs (left panel), 150 yrs (middle panel) and 300 yrs (right panel), and the results (dashed lines)
are all from 10 yrs.
To compare with the fiducial example (black lines), we fix the initial $ e_\star^0$ ($e_\IN^0$) in the upper (lower) panels (as labeled).
}
\label{fig:e-excitation}
\end{centering}
\end{figure*}

For coplanar ($\hat{\mathbf{L}}=\hat{\mathbf{L}}_\star$),
non-dissipative (no gravitational radiation) systems, the secular dynamics can be understood analytically.
When $e_\IN$, $e_\star\ll1$, the evolution of $\mathbf{e}_\IN$ and $\mathbf{e}_\star$ is governed
by the linear Laplace-Lagrange equations \cite{MD,YanqingWu}.
If we define the complex eccentricity variables as $\mathcal{E}_\IN\equiv e_\IN \mathrm{exp}(i\varpi_\IN)$ and
$\mathcal{E}_\star\equiv e_\star \mathrm{exp}(i\varpi_\star)$,
where $\varpi_\IN$, $\varpi_\star$ are the longitude of pericenter of the inner and outer orbits, then the evolution equations are reduced to
\begin{align}\label{eq:eigen Equation}
\frac{d}{dt} \begin{pmatrix} \mathcal{E}_\IN \\ \mathcal{E}_\star  \end{pmatrix} &= i \begin{pmatrix} \omega_\IN & \nu_{\IN,\star}
\\ \nu_{\star,\IN} & \omega_\star \end{pmatrix}
\begin{pmatrix} \mathcal{E}_\IN \\ \mathcal{E}_\star \end{pmatrix},
\end{align}
with
\begin{eqnarray}
&&\omega_\IN=\frac{3}{4}n_\IN\frac{m_\star}{m_{12}}\bigg(\frac{a_\IN}{a_\star}\bigg)^3+\omega_{\gr,\IN},\label{eq:A11}\\
&&\omega_\star=\frac{3}{4}n_\star\frac{m_1m_2}{m_{12}^2}\bigg(\frac{a_\IN}{a_\star}\bigg)^2+\omega_{\gr,\star}, \label{eq:A22}\\
&&\nu_{\IN,\star}=-\frac{15}{16}n_\IN\bigg(\frac{a_\IN}{a_\star}\bigg)^4\frac{m_\star(m_1-m_2)}{m_{12}^2},\label{eq:A12}\\
&&\nu_{\star,\IN}=-\frac{15}{16}n_\star\bigg(\frac{a_\IN}{a_\star}\bigg)^3\frac{m_1m_2(m_1-m_2)}{m_{12}^3},\label{eq:A21}
\end{eqnarray}
where $n_{\IN(\star)}=(G m_{12}/a_{\IN(\star)}^3)^{1/2}$ and
$\omega_{\gr,{\IN(\star)}}=3G^{3/2}m_{12}^{3/2}/[c^2a_{\IN(\star)}^{5/2}(1-e_{\IN(\star)}^2)]$ are
the mean motion and the GR-induced pericenter-precession frequency of the inner (outer) binary for $m_{12}\gg m_\star$, respectively.

Starting with $e_{\IN}=e_\IN^0$, $e_\star=0$ at $t=0$, Equation (\ref{eq:eigen Equation}) can be solved to determine the
time evolution of $e_\star(t)$ \citep[see][]{Liu-Yuan}, which oscillates between
$0$ and $e_\star^\m$, where
\be\label{eq:e max}
e_\star^\m=2e_\IN^0\frac{|\nu_{\star,\IN}|}{\sqrt{(\omega_\IN-\omega_\star)^2+4\nu_{\IN,\star}\nu_{\star,\IN}}}.
\ee
Clearly, $e_\star^\m$ attains its peak value when $\omega_\IN=\omega_\star$
occurs, and then we have
\ba\label{eq:e peak}
e^\mathrm{peak}_\star=e_\IN^0\frac{\mu^{1/2}_\IN}{\mu^{1/4}_\star m^{1/4}_\star}
\bigg(\frac{a_\IN}{a_\star}\bigg)^{1/4}.
\ea

\begin{figure*}
\begin{centering}
\includegraphics[width=5.5cm]{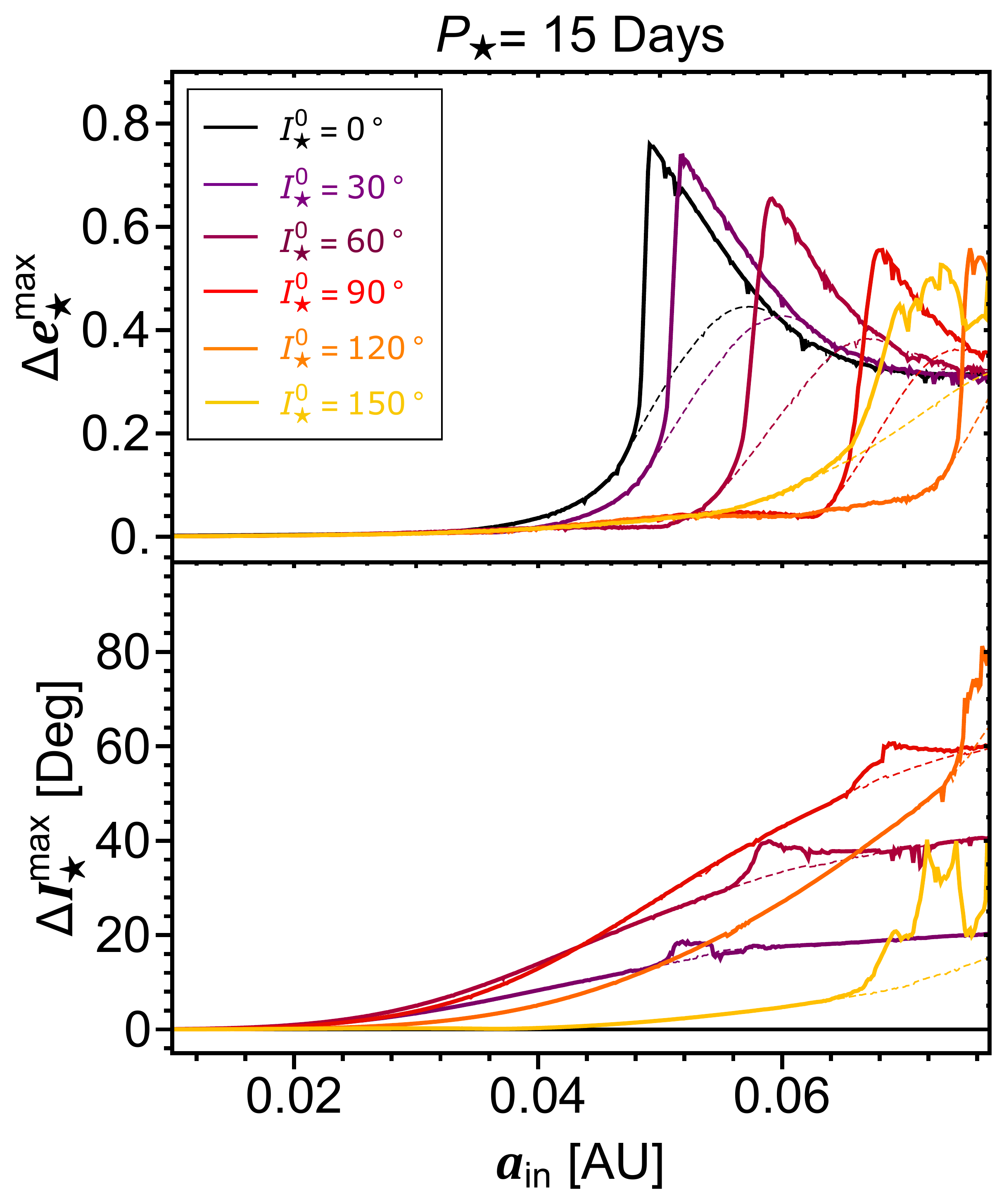}
\includegraphics[width=5.5cm]{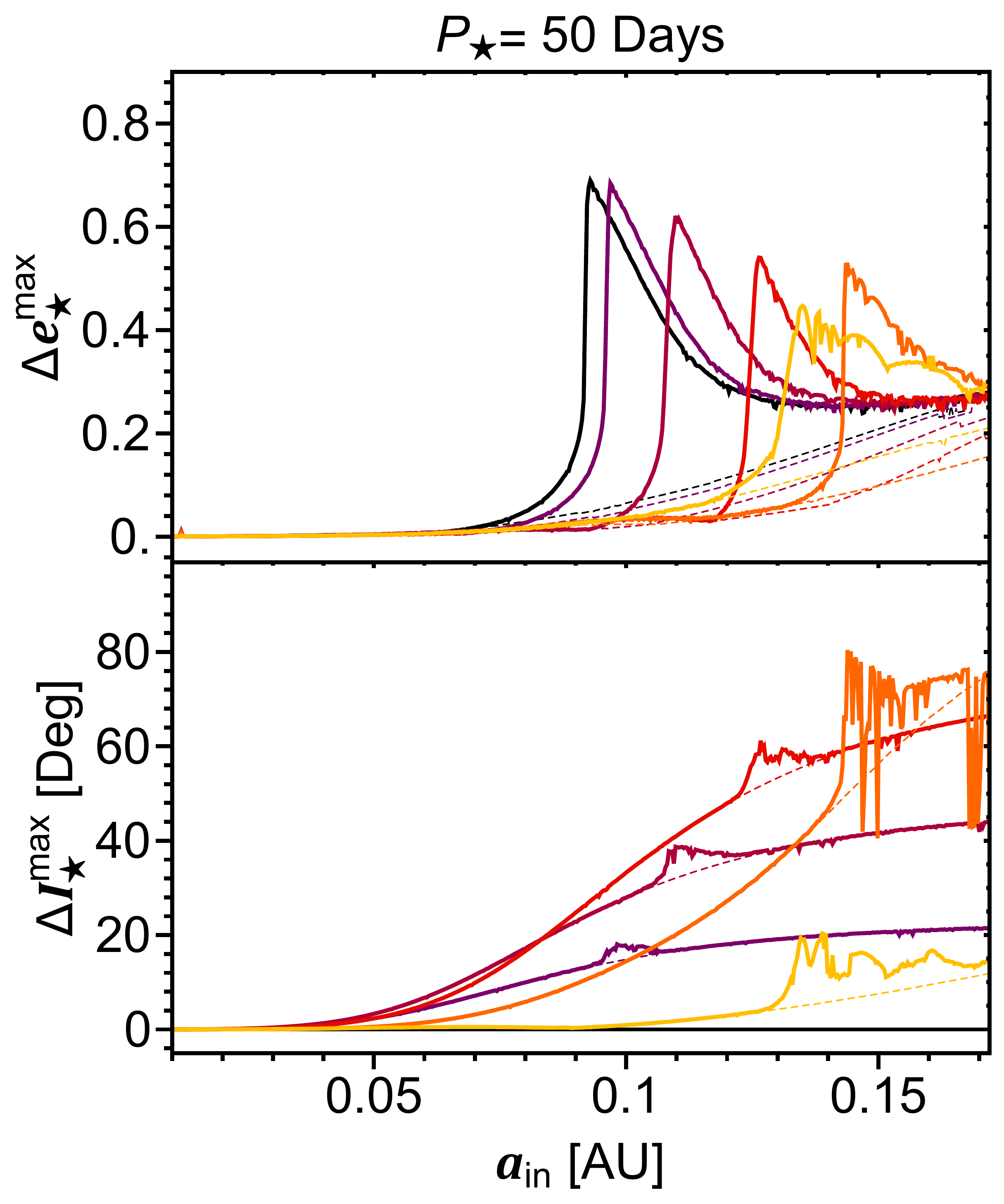}
\includegraphics[width=5.5cm]{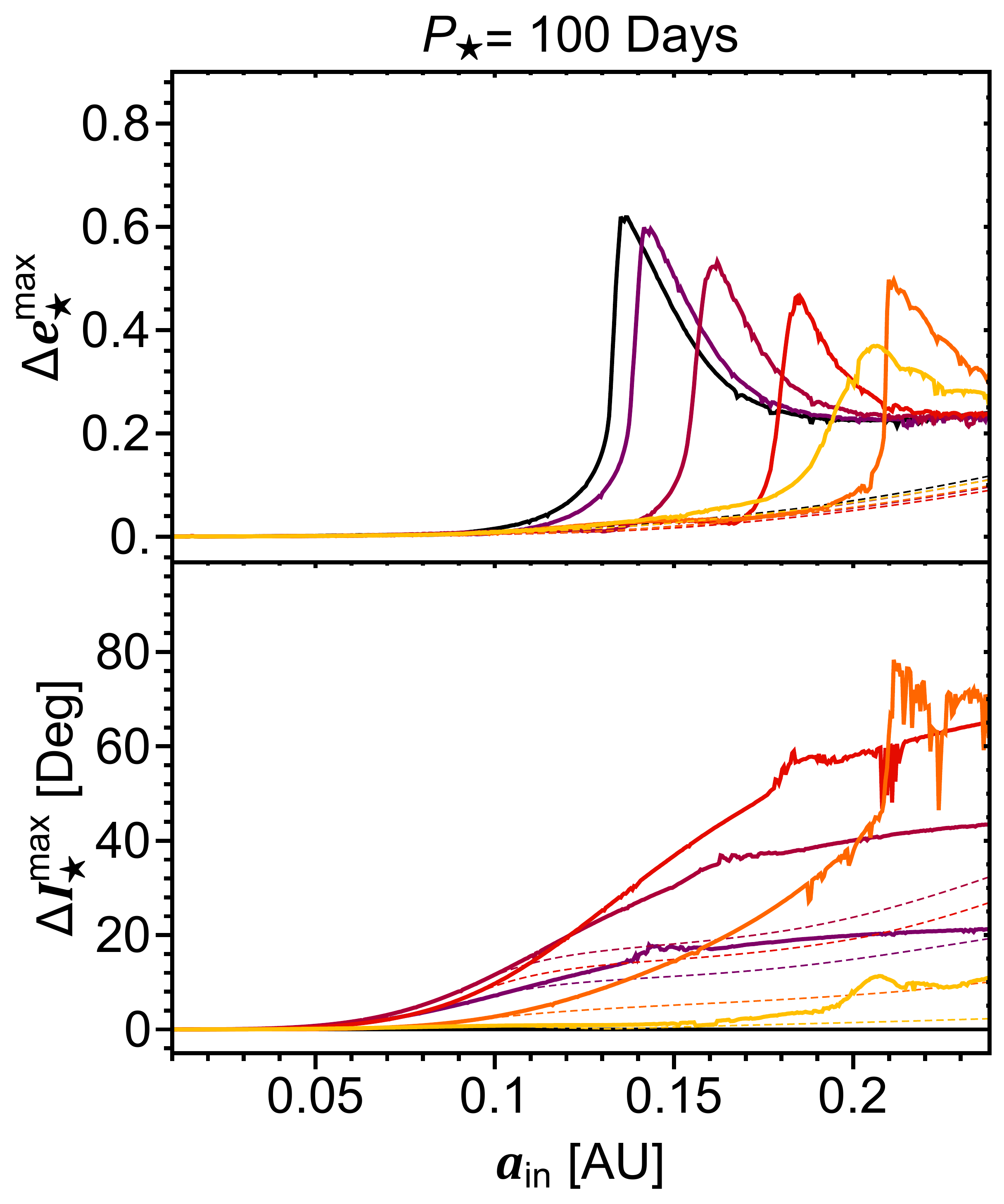}
\caption{
Similar to Figure \ref{fig:e-excitation},
but we set the initial eccentricities as $e_\IN^0=0.9$ and $e_\star^0=0$,
and consider different initial inclinations
(as labeled; $I_\star$ is the inclination angle between $\textbf{L}_\IN$ and $\textbf{L}_\star$).
The changes of $e_\star$ and $I_\star$ are shown in the upper and lower panels, respectively.
}
\label{fig:e-excitation 2}
\end{centering}
\end{figure*}

Note that the linear theory is valid for the low$-e$ systems.
For the example in Figure \ref{fig:Evolution},
non-zero $e_\IN^0$
may lead to unphysically large $e^\mathrm{peak}_\star$; however, Equations (\ref{eq:e max}) and (\ref{eq:e peak}) are useful
in the sense that we can expect:
\textit{i)} $e_\star-$excitation appears when $\omega_\IN\simeq\omega_\star$
(see the middle panel of Figure \ref{fig:Evolution}; the resonance occurs with finite $e_\star$);
\textit{ii)} $e_\star-$excitation becomes stronger with increasing $e_\IN$ (see also Figure \ref{fig:e-excitation} below).

For finite eccentricities, Equation (\ref{eq:eigen Equation}) breaks down.
However, in the case of exact coplanarity, the maximum eccentricity of a triple can still be calculated algebraically,
using energy and angular momentum conservations \cite{Lee}.
This method works well for two orbits with arbitrary eccentricities, but it cannot show the time evolution and resonance features.
A full derivation can be found in \cite{Liu-2020-PRD} and the solution is presented in Figure \ref{fig:Evolution}.

Figure \ref{fig:e-excitation} shows $e_\star-$excitation
with the arbitrary initial eccentricities for coplanar triples.
We again consider the fiducial example, in which the inner BHB has the total mass $m_{12}=50\ \rm{M}_\odot$ with mass ratio $m_2/m_1=0.2$.
We choose three values of the semimajor axis of the outer orbit ($a_\star$) and consider a range of $a_\IN$ that satisfies the stability criterion.
To illustrate the role of the eccentricity, we start with the same
initial configurations, namely, the argument of periapse, the longitude of ascending nodes
and the true anomaly of the outer orbit are set to be the same at $t=0$.
Each system is evolved for a long timescale (to achieve the highest value of $e_\star$) and a short timescale (10 yrs), respectively.
The maximum change of $e_\star$ is picked only for the system remains stable.

In the upper left panel of Figure \ref{fig:e-excitation}, we see that all the initial circular outer orbits can become eccentric
when the resonance occurs, and the maximum change $\Delta e_\star^\m$ grows as $e_\IN^0$ increases.
In the lower left panel, $e_\star-$excitation can still occur for the initially eccentric outer orbits.
However, due to the stability, only a fraction of systems (with a narrow range of $a_\IN$; when $e_\star^0\gtrsim0.6$)
may undergo significant eccentricity oscillations.
Note that for the wider stellar orbits ($P_\star=50, 100$ days), as shown in the middle and right panels,
the system has to be evolved for a sufficiently long time. This is because the timescale of $e_\star-$excitation
is of the order of \cite{Liuetal-2015}
\be\label{eq:Oct timescale}
T_{e_\star}\big|_{e_\IN,e_\star\ll1}\simeq \frac{m_{12}}{n_\IN\varepsilon_\oct}\bigg(\frac{a_\star}{a_\IN}\bigg)^3\frac{L_\star}{L_\IN}.
\ee

When the inner and outer orbits are mutually inclined, no simple analytical result can be derived,
and the long-term evolution of the outer orbits can only be studied numerically.
Previous studies \cite{Naoz-2017,Chiang-2018,Naoz-2019} showed that in inclined triple systems,
the eccentricity of the outer orbit can oscillate moderately and the angular momentum ($\textbf{L}_\star$)
undergoes nodal precession/nutation around the inner one ($\textbf{L}_\IN$).
Moreover, $\textbf{L}_\star$ might experience a flip if the tertiary is a test particle.

Figure \ref{fig:e-excitation 2} presents the results of the triples with a series of initial inclinations.
We see that in the upper panels, regardless of the values of the initial inclinations,
$e_\star-$excitation due to resonance always occurs, and the resonance location
shifts when $I_\star^0$ changes ($I_\star^0$ is the initial inclination angle between $\textbf{L}_\IN$ and $\textbf{L}_\star$).
In the lower panels,
we find that the inclination varies for a wider range of $a_\IN$ compared to the change of $e_\star$.
In particular, $\Delta I_\star$ always undergoes an additional excitation when $\Delta e_\star^\m$ approaches the peak value
for the inclined systems.

\section{Resonance in stellar orbits and detectability}

We now focus on dependence of the change in eccentricity on parameters of the outer stellar orbit, considering the inner BHB is a LISA source.
To explore the observability of this effect,
we consider the detectability of a maximum eccentricity change, $\Delta e_\star^\mathrm{peak}$, within a certain timescale.

We initialize systems with $e_\IN^0=0.9$ and $e_\star^0=0$ in a coplanar configuration.
For the inner BHB, we choose three values of the total mass ($m_{12}/\rm{M_{\odot}}=20,50,100$),
and allow the mass-ratio range to take on all values such that both masses are consistent with being a BH, $m_1, m_2\gtrsim5\rm{M}_\odot$.
To develop the joint detection with GW detectors in the coming future,
we focus on the BHBs radiating GWs in the LISA frequency band.
Thus, the semimajor axis $a_\IN$ is chosen from a uniform distribution that satisfies $f_\gw\geq10^{-4}$ Hz
and with merger time (due to GW emission) larger than $10^3$ yrs \cite{Peters-1964}.
Then, for the (outer) stellar orbits, we sample the semimajor axis $a_\star$ from a range of
$P_\star\sim[1, 180]$ days, considering only systems that are dynamically stable.
Each system evolves to 10, 30 or 100 yrs, using the SA equations of motion.
The maximum change of the $e_\star$ is recorded if the star remains gravitationally bound and stable during the evolution.
Finally, since the orbital evolution relies on the initial geometry of the triples,
to cover all possibilities, we randomly sample the longitude of pericenter of the inner orbit and the true anomaly of the outer orbit.
Each triple system is evolved with 100 different initial geometries.

\begin{figure}
\begin{centering}
\includegraphics[width=7cm]{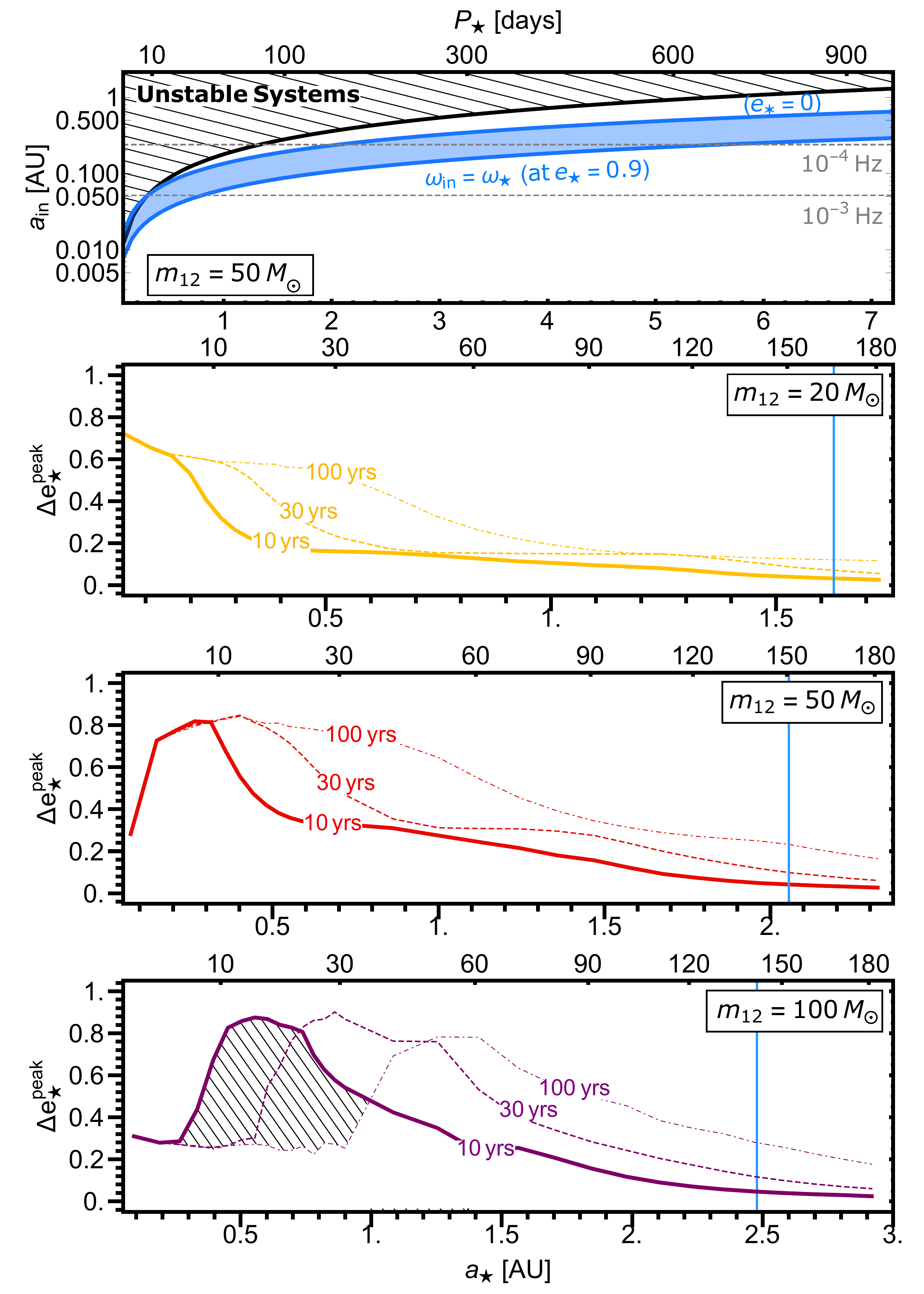}
\caption{Top panel: Parameter space in [$a_\IN-a_\star$ or $(P_\star)$] plane, where the apsidal precession resonance occurs.
We consider the BHB with $m_{12}=50\ \rm{M}_\odot$, $m_2/m_1=0.2$ and the initial $e_\IN^0=0.9$, $e_\star^0=0$.
The blue region is given by Equations \ref{eq:A11}-\ref{eq:A22} including the finite $e_\IN^0$ and $e_\star$ (as labeled).
The dashed lines characterize the frequency of GWs emitted by the inner BHB.
Lower panels: Maximum of $\Delta e_\star^\m$ (i.e., $\Delta e_\star^\mathrm{peak}$) as a function of $a_\star$ for different $m_{12}$.
The solid and dashed lines are obtained by the numerical integrations for 10, 30 and 100 yrs, respectively.
The vertical blue line refers to $a_\star=a_\star^\mathrm{Res}$ at $e_\star=0$ via $f_\gw=10^{-4}$ Hz.
The cross-hatched region in the bottom panel indicates that the systems eventually become unbound when evolving for a longer timescale.
}
\label{fig:Different Pout}
\end{centering}
\end{figure}

For a given set of parameters, the criterion of apsidal precession resonance ($\omega_\IN=\omega_\star$)
provides a good estimate for the resonance radius:
resonance occurs at the location $a_\star=a_\star^\mathrm{Res}$ for a given $a_\IN$.
In Figure \ref{fig:Different Pout}, the upper panel clarifies the resonance locations when $m_{12}=50\rm{M}_\odot$.
The region of interest where the outer orbit potentially undergoing
$e_\star-$excitation due to the apsidal precession resonance is located within a wide range of $a_\star$ (or $P_\star$).

The lower panels of Figure \ref{fig:Different Pout} present the results of the averaged maximum changes of the outer eccentricity
($\Delta e_\star^\mathrm{peak}$),
over all 100 runs, as a function of $a_\star$ for different masses of BHBs.
A remarkable eccentricity excitation is achieved for all tested values of the BHB mass. For the $m_{12}=20\rm{M}_\odot$ case,
the maximum eccentricity growth is found for the smallest $a_\star$ and decreases for larger $a_\star$.
For the BHBs ($m_{12}/\rm{M_{\odot}}=50,100$),
the perturbation becomes stronger,
and the induced peak eccentricity can be so high (because of $e^\mathrm{peak}_\star\varpropto\mu_\IN$)
that some of the systems with small $a_\star$ can become unbound.
The peak of $\Delta e_\star^\mathrm{peak}$ appears to shift to larger $a_\star$ for the larger BHB masses.
The decrease of $\Delta e^\mathrm{peak}$ with increasing $a_\star$ is the result of the secular timescale of generating the
eccentricity growth (See Figure \ref{fig:e-excitation 2}).
Note that, in principle, the resonance can occur as $a_\star\leq a_\star^\mathrm{Res}$ (the vertical blue line; $P_\star\lesssim 160$ days).
As shown, more systems can have larger $\Delta e_\star^\mathrm{peak}$
when the evolving time is longer (See Figure \ref{fig:e-excitation}).
We find that $e_\star-$excitation is mainly contributed by the inner BHBs with relatively small mass ratios ($m_2/m_1\lesssim0.5$).

The secular variability of the orbital eccentricity induces
a change in the projected orbit that may be probed via astrometric monitoring with surveys such as \textit{Gaia} \cite{Gaia}.
To determine detectability with \textit{Gaia},
we compute a signal-to-noise ratio (SNR) for astrometric detection,
$\rho=\theta_\mathrm{signal}/\theta_\mathrm{Gaia}$.
Assuming the signal can be well-approximated by the change in apocenter, the maximum such signal for a system at distance $D$ is approximated by
\be
\label{eq:Source resolution}
\theta_\mathrm{signal}(D)\simeq\frac{a_\star\Delta e_\star^\mathrm{peak}}{D}.
\ee

To compute noise, we follow \cite{Dan-Gaia}, which draws on \cite{Casertano-1995,Bernstein,Casertano-2008,Perryman,Ranalli}
to evaluate the \textit{Gaia} astrometric precision for a single-scan.
Here, we assume that our source is a solar-type star with absolute V-band magnitude of $-26.8$,
and $V-I_c=0.688$ \cite{Holmberg}. Following \citep{Perryman}, the single-scan precision is computed from the end-of-mission,
sky-averaged parallax uncertainty $\theta_{\mathrm{eom}}(D)$, as,
\be
\label{eq:Gaia resolution}
\theta_\mathrm{Gaia}(D) \equiv  \frac{\sqrt{140}}{1.1 \times 2.15 } \theta_{\mathrm{eom}}(D).
\ee
The pre-factors account for an average 140 \textit{Gaia} visits over 10 years,
a geometrical averaging factor of $2.15$, and a contingency margin of $1.1$ \cite{Perryman}.
Note that the contingency margin is 1.1 instead of the value of 1.2 chosen in \citep{Perryman}.
This is based on newest information from Gaia EDR3 uncertainties as expressed in Section 1 of \citep{GaiaFactSheet}.
To compute the end-of-mission astrometric precision,
we use the most up-to-date fitting formula from the Gaia expected science performance document \citep{GaiaFactSheet},
\ba
\label{eq:Gaia Resolution}
\theta_{\mathrm{eom}}(D) &=&  0.527 \left[  40 + 800 Z  + 30 Z^2  \right]^{1/2}  \mu \mathrm{as}  \\ \nonumber
Z &\equiv& \max\left\{  10^{0.4\left[13.0-15.0\right] } ,  10^{0.4\left[G-15.0\right] } \right\} \\ \nonumber
G &=& m_V(D) - 0.01746 + 0.008092(V - I_C)\\ \nonumber
&&- 0.2810 (V - I_C)^2 + 0.03655 (V - I_C)^3.
\ea
Here the $0.527$ prefactor is for a 10 year mission (referred to as Gaia DR5 in \cite{GaiaFactSheet}),
the conversion between Gaia-G- and V-magnitudes,
in the last line above, is given in Table A2 of \citep{Evans+2018}, and $m_V(D) = -26.8 + 2.5\log_{10}[(D/\au)^2]$
is the apparent magnitude of a sun-like star at distance $D$ in Astronomical Units (AU).

\begin{figure}
\begin{centering}
\includegraphics[width=7cm]{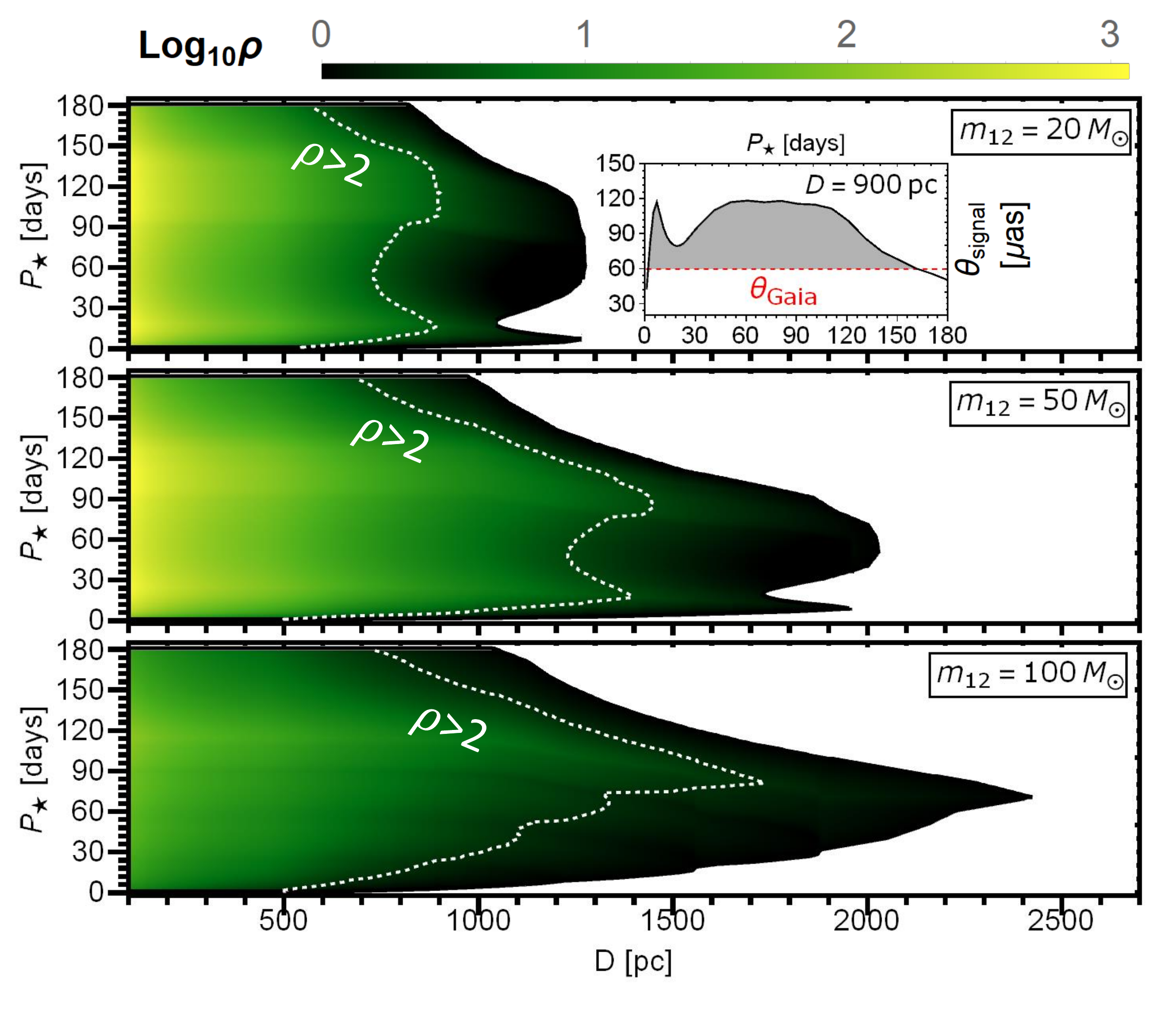}
\caption{Detectability of the change of the outer stellar eccentricity in the ($P_\star-D$) plane with \textit{Gaia},
which is evaluated by Equations (\ref{eq:Source resolution})-(\ref{eq:Gaia resolution})
and $\Delta e_\star^\mathrm{peak}$ is given by the data from Figure \ref{fig:Different Pout} for 10 yrs.
The subfigure in the top panel illustrates an example of detectability where the source is
at $D=900$ pc,
and the shaded region corresponds to the systems with $\theta_\mathrm{signal}\geq\theta_\mathrm{Gaia}$.
The white dashed lines specify the region with $\rho>2$.
}
\label{fig:Detectability}
\end{centering}
\end{figure}

Figure \ref{fig:Detectability} summarizes
the optimal-characteristic SNR as a function of distance to the source and outer orbital period,
for three different inner binary masses.
Here, we use the data from Figure \ref{fig:Different Pout}
that gives the maximum change of eccentricity
for a range of $a_\star$ over 10 yrs
(for $m_{12}=100\ \rm{M}_\odot$, we assume $\Delta e_\star^\mathrm{peak}\lesssim0.3$ due to
the long term instability).
We see that the overall SNR improves as the distance $D$ decreases, and
the boundary of detectability is marked at $\sim10^3$ pc
given by $\rho=2$ \citep{Ranalli}.
Two peaks of high SNR are the results of the
pure $e_\star-$excitation (Figure \ref{fig:Different Pout})
and the enhancement of $\theta_\mathrm{signal}$ from wide binaries.
Although the detectability here is evaluated based only on $\Delta e_\star^\mathrm{peak}$,
the outer pericenter argument can vary in the actual detection \cite{Suto-2}.
When the triple system is inclined,
a combination of predicted changes in orbital inclination
and line-of-sight orbital projection can also increase or decrease our estimate here.
Our results are characteristic of the optimal astrometric signatures over the course of the \textit{Gaia} mission.

Note that the average time between visits is $\sim 26$ days,
hence, the shorter period systems
may be sampled at sub-orbital frequencies.
However, because this is a secular effect,
even a longer than orbital sampling rate could probe the longer-timescale secular change of the projected orbit.
Follow-up analysis must simulate mock systems to determine the required orbital sampling rate
and SNR needed to reliably detect this effect.
Further work is also required to determine
in which cases the current \textit{Gaia} pipeline would flag such secularly evolving systems as binaries with an unseen companion,
or miss identify them due to difficulty in fitting to a Keplerian orbit.

\section{Summary and Discussion}

We have studied a novel secular dynamical effect of a solar-type star around a compact BHB in a nearly coplanar
triple configuration. We point out that the stellar orbit may undergo significant oscillations of the orbital eccentricity
if the system satisfies an ``apsidal precession resonance''.
Starting with an eccentric inner BHB, the outer eccentricity can be excited due to the resonance and
the enhancement becomes stronger as the inner eccentricity increases.
This effect, which can drive the eccentricity of the outer orbits close to unity (even unbound the binary), is overlooked in all previous studies and
may have applications for other types of systems, such as the planet around a binary star
and star/compact object around a supermassive BH binary with/without a gaseous disc.

\begin{figure}
\begin{centering}
\includegraphics[width=4.55cm]{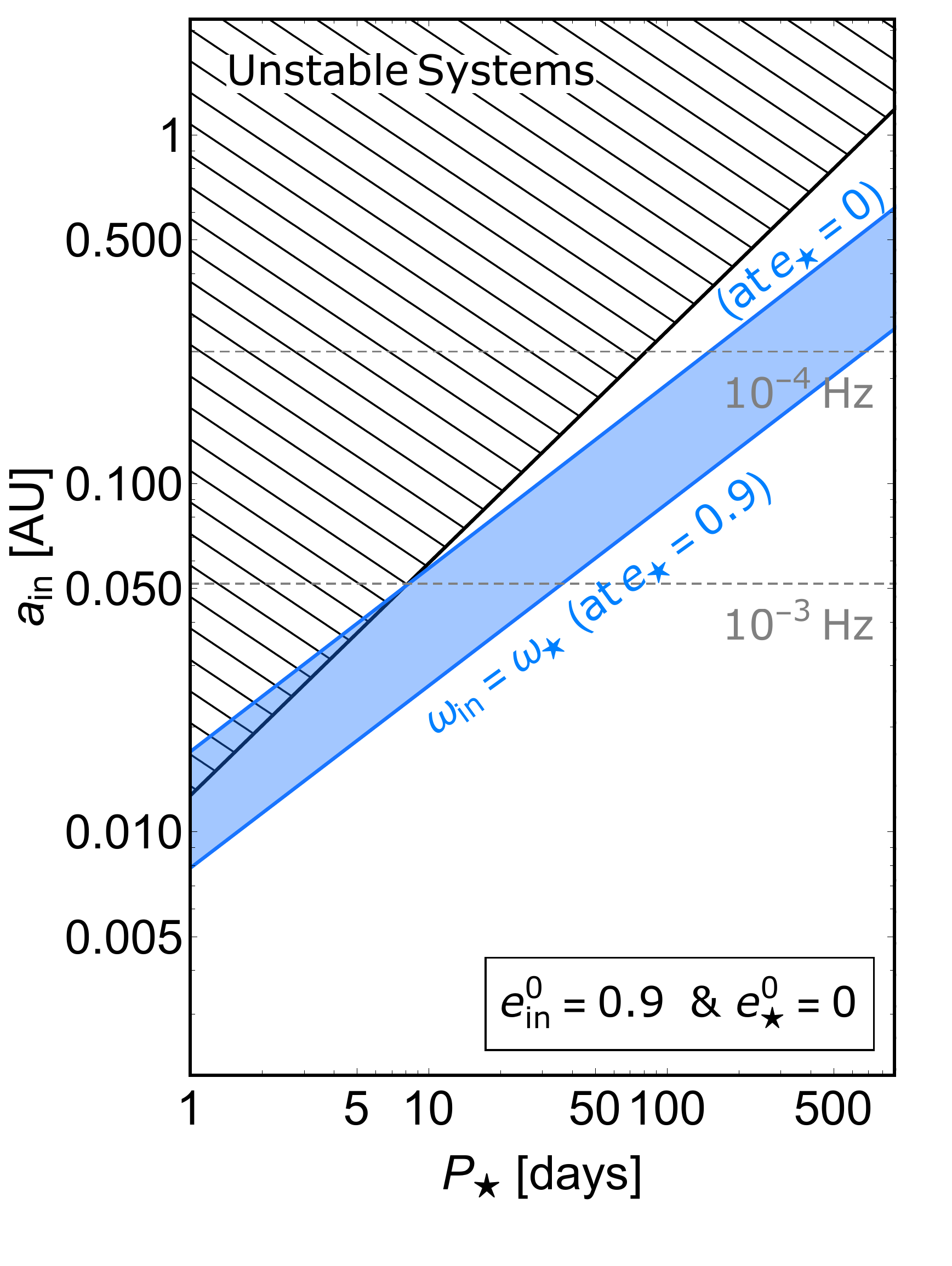}
\includegraphics[width=3.7cm]{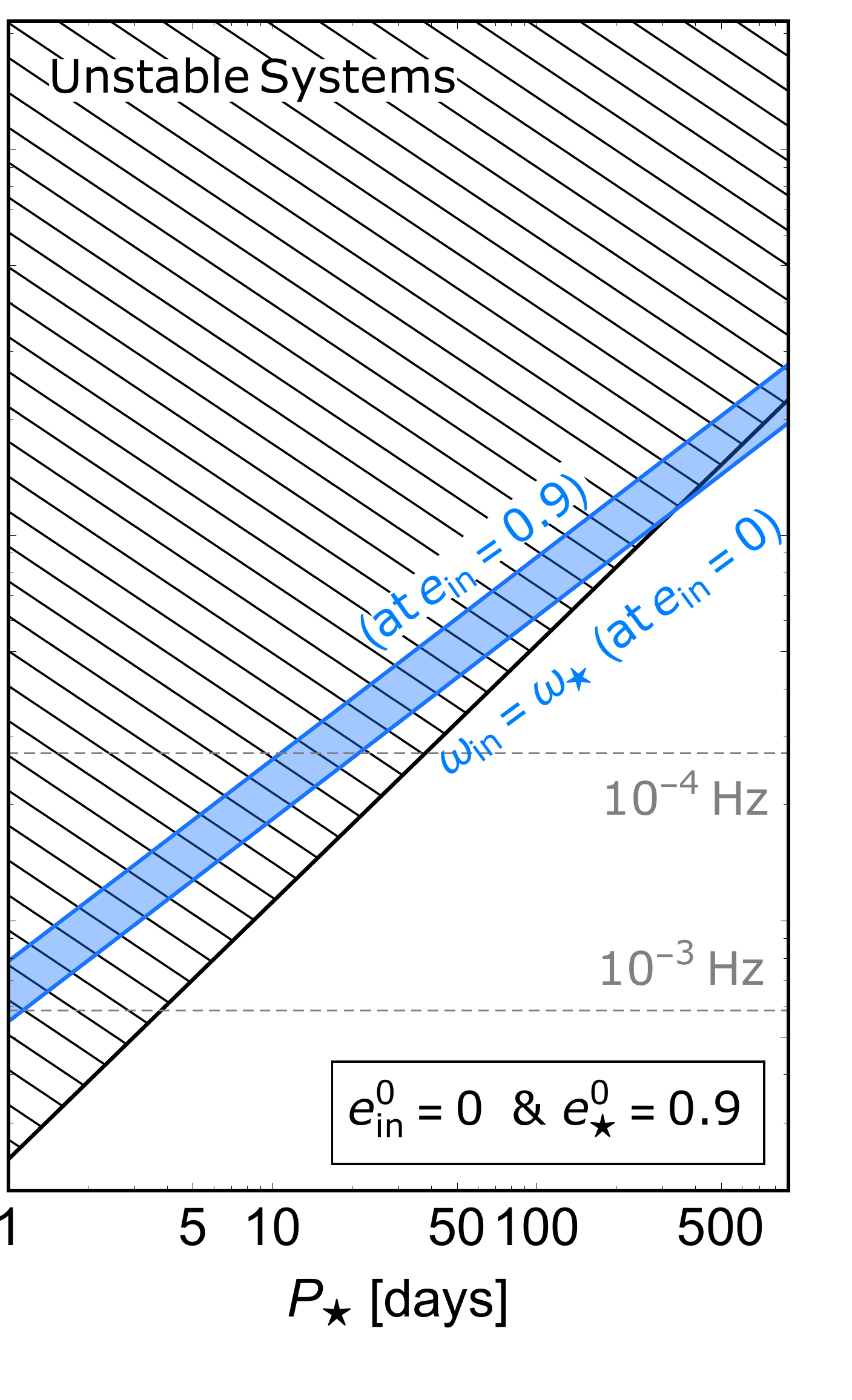}
\caption{Similar to the top panel of Figure \ref{fig:Different Pout},
but we include two combinations of initial eccentricities (as labeled).
Here, the mass of BHB is set to be $m_{12}=50\ \rm{M}_\odot$ and the mass ratio is $m_2/m_1=0.2$.
}
\label{fig:parameter space}
\end{centering}
\end{figure}

\begin{figure}
\begin{centering}
\includegraphics[width=4.62cm]{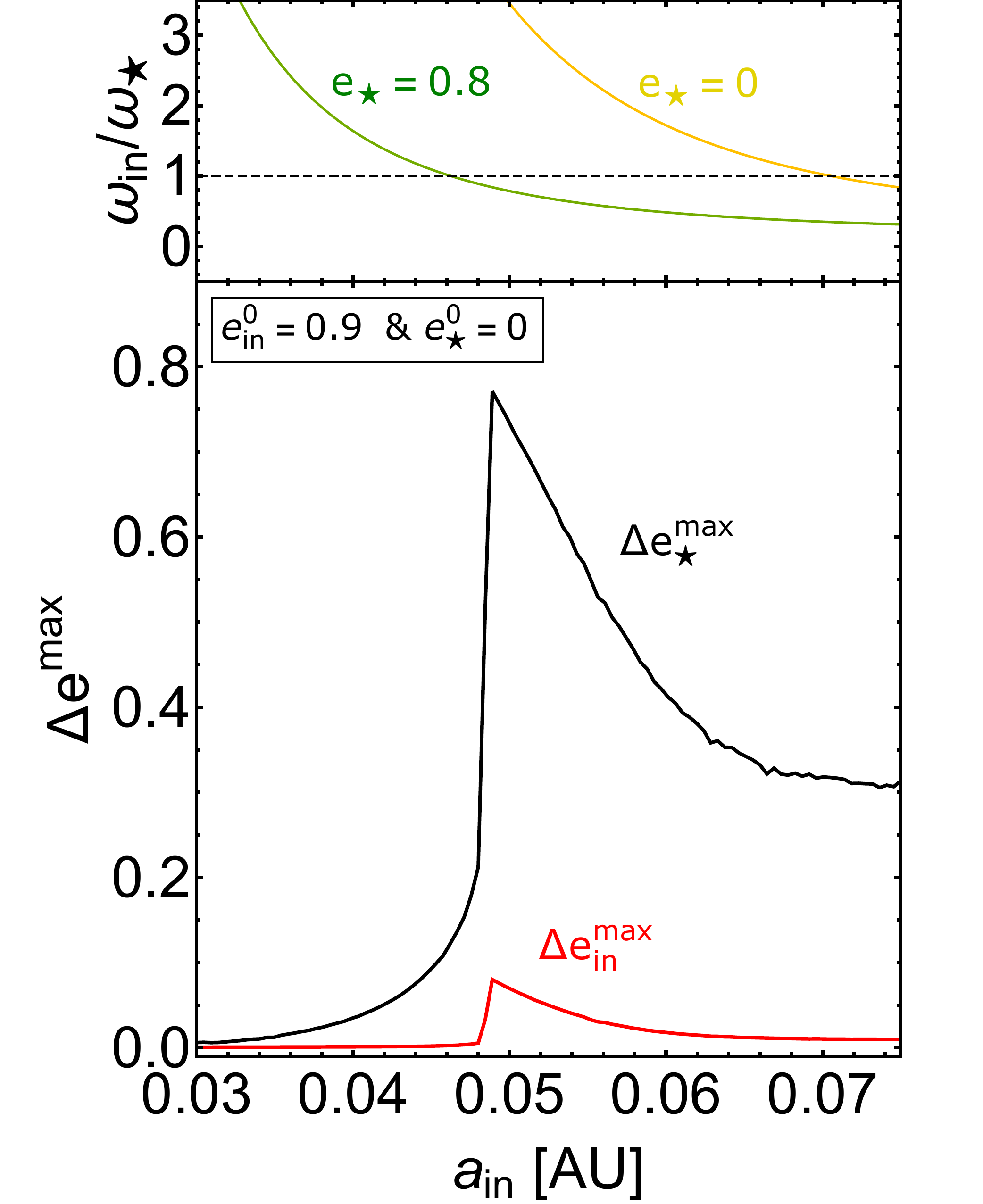}
\includegraphics[width=3.7cm]{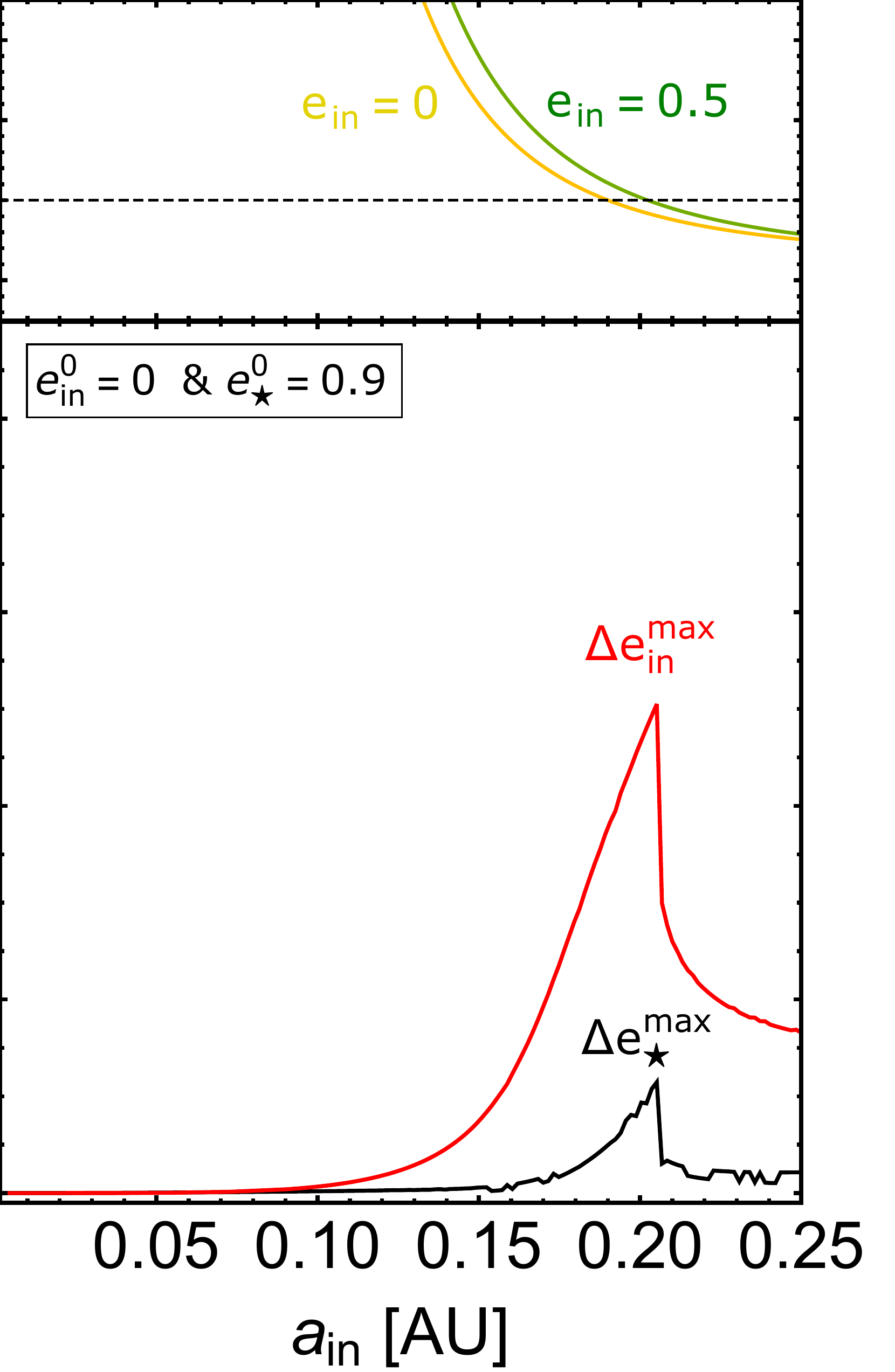}
\caption{Similar to the middle panel of Figure \ref{fig:Evolution}, but we include the change of eccentricities
of both inner and outer binaries.
The system parameters are $m_{12}=50\ \rm{M}_\odot$, $m_2/m_1=0.2$,
$e_\IN^0=0.9$, $e_\star^0=0$, $P_\star=15$ days (left panel),
$e_\IN^0=0$, $e_\star^0=0.9$, $P_\star=800$ days (right panel).
Upper panels: the ratio of apsidal precession rate.
Lower panels: the results (black and red lines) are obtained by the numerical integrations
of SA equations for 25 yrs (left panel) and $6\times10^3$ yrs (right panel).
}
\label{fig:e-excitation 3}
\end{centering}
\end{figure}

Note that the apsidal precession resonance allows
angular momentum exchange to occur efficiently between the inner and outer orbits, leading to
a transfer of eccentricity. One case that we do not include in our study is
the triple systems with $e_\IN^0=0$ and $e_\star^0=0.9$.
In this situation, the inner (outer) binary is expected to become (less) eccentric as the resonance occurs.

Figure \ref{fig:parameter space} shows the parameter space where the apsidal precession resonance can play a role,
on the basis of $\omega_\IN (e_\IN)=\omega_\star (e_\star)$.
Compared to the fiducial cases (left panel), the resonance region for the initially stable systems in the right panel
is shifted to larger $a_\star$ (or $P_\star$).
Therefore, the corresponding semimajor axis of the inner BHB ($a_\IN$) becomes larger,
leading to a much lower GW frequency range (i.e., out of the LISA band).

Figure \ref{fig:e-excitation 3} presents the change of eccentricities
due to resonance for both inner and outer orbits,
taking into account the examples identified in the Figure \ref{fig:parameter space}.
In the left panel (fiducial case),
we see that
the inner eccentricity decreases (i.e., $\Delta e_\IN^\m\lesssim0.1$) when the outer eccentricity is excited.
In the right panel, we find that the eccentricities evolve in an opposite way.
The inner binary can efficiently gain some eccentricity from the outer binary during the resonance
(the peak value is about $\Delta e_\IN^\m\sim0.5$),
while the the outer eccentricity can decrease by a factor of $\lesssim0.2$.

Since we are interested in the LISA source and the evidence change of $e_\star$ within $\sim10$ years,
we do not include this type of system (with $e_\IN^0=0$ and $e_\star^0=0.9$) in our main study.

The formation of compact massive BHB$+$star systems may be challenging.
The progenitor stars of these BHBs usually expand hundreds or thousand of solar radii throughout their evolution,
likely dynamically interacting with the stellar tertiary.
However, some low-metallicity massive stellar binaries might remain compact
throughout their evolution \cite{Marchant-2016}, allowing for dynamically-stable compact triples \cite{Alejandro2021}.
Alternatively, a BHB can be formed first and eventually capture a long-lived low-mass star.
Note that our analysis is not restricted to any specific formation scenarios, and can be adapted to other types of systems by applying scaling relations.

We find that the secular variability of the stellar orbit's apocenter induced by the changing eccentricity
is detectable by \textit{Gaia}
for inner BHBs emitting GWs in the LISA frequency range.
Assuming that the formation and merger rates of BHBs are in equilibrium,
we expect to have hundreds of BHBs in the LISA band in the Galaxy based on the LIGO detection rate
\cite{XiaoFang,Christian,Wagg,Kremer,Xianyu-LISA,Hoang}.
Since our proposed secular variability can only be resolved by \textit{Gaia} within several kpc,
the expected number of sources that LISA and \textit{Gaia} could see becomes $\sim$a few.
Although the actual number of LISA sources accompanied by a tertiary star in our Galaxy is quite uncertain, identifying the secular motion of stellar orbits in current (Gaia Data Release 3; \cite{Gaia}) and future \textit{Gaia} data is timely.

Our proof-of-concept calculations demonstrate that the long-term evolution of eccentricity of a nearby stellar orbit
can serve as a distinctive imprint of such an unseen binary companion.
Precise measurements of secular variability are therefore an independent approach to
reveal hidden BHBs, in addition to GW detection.
We emphasize that the inner binary systems which generate \textit{Gaia}-detectable variations
in the orbits of their stellar tertiary also emit GWs in the LISA band.
Therefore, \textit{Gaia} may provide
candidates LISA sources before LISA flies (planned for the 2030's).
In this sense,
a joint detection with \textit{Gaia} and LISA \cite{Silvia-LISA,Dan-LISA,Breivik}
would be a unique multi-messenger tool to understand the evolution, fate and configurations of compact BHBs.

\textit{Acknowledgement.---} We thank Dong Lai and Bence Kocsis for useful discussions.
We also thank Hagai Perets and Jihad Touma for careful reading the manuscript and helpful comments given.
B.L. gratefully acknowledges support from
the European Union's Horizon 2021 research and innovation programme under the Marie Sklodowska-Curie grant agreement No. 101065374.
D.J.D. received funding from the European Union's Horizon 2020
research and innovation programme under Marie Sklodowska-Curie grant agreement No. 101029157,
and from the Danish Independent Research Fund through Sapere Aude Starting Grant No. 121587.
A.V.G. received support through Villum Fonden grant No. 29466.
J.S. is supported by the Villum Fonden grant No. 29466.


\begin{thebibliography}{100}

\bibitem{LIGO-2021} R. Abbott et al. (LIGO Scientific, VIRGO), arXiv:2108.01045 [gr-qc] (2021).

\bibitem{Lipunov-1997} V. M. Lipunov, K. A. Postnov, and M. E. Prokhorov, Astron. Lett. {\bf 23}, 492 (1997).

\bibitem{Lipunov-2007} V. M. Lipunov, V. Kornilov, E. Gorbovskoy, D. A. H. Buckley, N. Tiurina, P. Balanutsa, A. Kuznetsov, J. Greiner,
V. Vladimirov, D. Vlasenko et al., Mon. Not. R. Astron. Soc. {\bf 465}, 3656 (2017).

\bibitem{Podsiadlowski-2003} P. Podsiadlowski, S. Rappaport, and Z. Han, Mon. Not. R. Astron. Soc. {\bf 341}, 385 (2003).

\bibitem{Belczynski-2010} K. Belczynski, M. Dominik, T. Bulik, R. O'Shaughnessy,
C. Fryer, and D. E. Holz, Astrophys. J. Lett. {\bf 715}, L138 (2010).

\bibitem{Belczynski-2016} K. Belczynski, D. E. Holz, T. Bulik, and R. O'Shaughnessy, Nature (London) {\bf 534}, 512 (2016).

\bibitem{Dominik-2012} M. Dominik, K. Belczynski, C. Fryer, D. E. Holz, E. Berti, T. Bulik, I. Mandel, and R. O'Shaughnessy,
Astrophys. J. {\bf 759}, 52 (2012).

\bibitem{Dominik-2013} M. Dominik, K. Belczynski, C. Fryer, D. E. Holz, E. Berti, T. Bulik, I. Mandel, and R. O'Shaughnessy,
Astrophys. J. {\bf 779}, 72 (2013).

\bibitem{Dominik-2015} M. Dominik, E. Berti, R. O'Shaughnessy, I. Mandel, K. Belczynski, C. Fryer, D. E. Holz, T. Bulik, and F. Pannarale,
Astrophys. J. {\bf 806}, 263 (2015).

\bibitem{Alejandro-2017} S. Stevenson, A. Vigna-G\'omez, I. Mandel, J. W. Barrett, C. J. Neijssel, D. Perkins, and
S. E. de Mink, Nature Commun. {\bf 8}, 14906 (2017).

\bibitem{Mandel-2016} I. Mandel and S. E. De Mink, Mon. Not. R. Astron. Soc. {\bf 458}, 2634 (2016).

\bibitem{Marchant-2016} P. Marchant, N. Langer, P. Podsiadlowski, T. M. Tauris, and T. J. Moriya, Astron. Astrophys. {\bf 588}, A50 (2016).

\bibitem{duBuisson} L. du Buisson, P. Marchant, P. Podsiadlowski, C. Kobayashi, F. B. Abdalla, P. Taylor, I. Mandel,
S. E. de Mink, T. J. Moriya, and N. Langer, Mon. Not. R. Astron. Soc. {\bf 499}, 5941 (2020).

\bibitem{Riley} J. Riley, I. Mandel, P. Marchant, E. Butler, K. Nathaniel, C. Neijssel, Shortt S., and A. Vigna-G\'omez,
Mon. Not. R. Astron. Soc. {\bf 505}, 663 (2021).

\bibitem{Baruteau-2011} C. Baruteau, J. Cuadra, and D. N. C. Lin, Astrophys. J. {\bf 726}, 28 (2011).

\bibitem{McKernan-2012} B. McKernan, K. E. S. Ford, W. Lyra, and H. B. Perets, Mon. Not. Roy. Astron. Soc. {\bf 425}, 460 (2012).

\bibitem{McKernan-2018} B. McKernan, K. E. S. Ford, J. Bellovary, N. W. C. Leigh, Z. Haiman, B. Kocsis, W. Lyra, M. M. MacLow, B. Metzger, M. O'Dowd,
S. Endlich, and D. J. Rosen, Astrophys. J. {\bf 866}, 66 (2018).

\bibitem{Bartos-2017} I. Bartos, B. Kocsis, Z. Haiman, and S. M{\'a}rka, Astrophys. J. {\bf 835}, 165 (2017).

\bibitem{Stone-2017} N. C. Stone, B. D. Metzger, and Z. Haiman, Mon. Not. Roy. Astron. Soc. {\bf 464}, 946 (2017).

\bibitem{Leigh-2018} N. W. C. Leigh, A. M. Geller, B. McKernan, K. E. S. Ford, M. M. Mac Low, J. Bellovary, Z. Haiman, W. Lyra,
J. Samsing, M. O’Dowd, B. Kocsis, and S. Endlich,  Mon. Not. Roy. Astron. Soc. {\bf 474}, 5672 (2018).

\bibitem{Secunda-2019} A. Secunda, J. Bellovary, M.-M. Mac Low, K. E. S. Ford, B. McKernan, N. W. C. Leigh, W. Lyra, and Z. S{\'a}ndor,
Astrophys. J. {\bf 878}, 85 (2019).

\bibitem{Yang-2019} Y. Yang, I. Bartos, V. Gayathri, K. E. S. Ford, Z. Haiman, S. Klimenko, B. Kocsis, S. M{\'a}rka, Z. M{\'a}rka,
B. McKernan, and R. O'Shaughnessy, Phys. Rev. Lett. {\bf 123}, 181101 (2019).

\bibitem{Grobner-2020} M. Gr\"{o}bner, W. Ishibashi, S. Tiwari, M. Haney, and P. Jetzer, Astron. Astrophys. {\bf 638}, A119 (2020).

\bibitem{Ishibashi-2020}W. Ishibashi and M. Gr\"{o}bner, Astron. Astrophys. {\bf 639}, A108 (2020).

\bibitem{Tagawa-2020} H. Tagawa, Z. Haiman, and B. Kocsis, Astrophys. J. {\bf 898}, 25 (2020).

\bibitem{Liyaping-2021} Y. P. Li, A.~M. Dempsey, S. Li, H. Li, and J. Li, Astrophys. J. {\bf 911}, 124 (2021).

\bibitem{Ford-2021} K. E. S. Ford and B. McKernan, arXiv:2109.03212 [astro-ph.HE] (2021).

\bibitem{Samsing-Nature} J. Samsing, I. Bartos, D.~J. D'Orazio, Z. Haiman, B. Kocsis, N.~W.~C. Leigh, B. Liu, M.~E. Pessah, and H. Tagawa,
Nature, {\bf 603}, 237 (2022).

\bibitem{Lirixin-2022} R. Li and D. Lai, arXiv:2202.07633 [astro-ph.HE] (2022).

\bibitem{Lijiaru-2022} J. Li and D. Lai, arXiv:2203.05584 [astro-ph.HE] (2022).

\bibitem{Zwart(2000)} S. F. P. Zwart and S. L. W. McMillan, Astrophys. J. Lett. {\bf 528}, L17 (2000).

\bibitem{OLeary(2006)} R. M. O'Leary, F. A. Rasio, J. M. Fregeau, N. Ivanova, and R. OShaughnessy, Astrophys. J. {\bf 637}, 937 (2006).

\bibitem{Miller(2009)} M. C. Miller and V. Lauburg, Astrophys. J. {\bf 692}, 917 (2009).

\bibitem{Banerjee(2010)} S. Banerjee, H. Baumgardt, and P. Kroupa, Mon. Not. R. Astron. Soc. {\bf 402}, 371 (2010).

\bibitem{Downing(2010)} J. M. B. Downing, M. J. Benacquista, M. Giersz, and R. Spurzem, Mon. Not. R. Astron. Soc. {\bf 407}, 1946 (2010).

\bibitem{Ziosi(2014)} B. M. Ziosi, M. Mapelli, M. Branchesi, and G. Tormen, Mon. Not. R. Astron. Soc. {\bf 441}, 3703 (2014).

\bibitem{Rodriguez(2015)} C. L. Rodriguez, M. Morscher, B. Pattabiraman, S. Chatterjee, C.-J. Haster, and F. A. Rasio,
Phys. Rev. Lett. {\bf 115}, 051101 (2015).

\bibitem{Samsing(2017)} J. Samsing and E. Ramirez-Ruiz, Astrophys. J. Lett. {\bf 840}, L14 (2017).

\bibitem{Samsing(2018)} J. Samsing and D. J. D'Orazio, Mon. Not. R. Astron. Soc. {\bf 481}, 5445 (2018).

\bibitem{Rodriguez(2018)} C. L. Rodriguez, P. Amaro-Seoane, S. Chatterjee, and F. A. Rasio, Phys. Rev. Lett. {\bf 120}, 151101 (2018).

\bibitem{Gondan(2018)} L. Gond\'{a}n, B. Kocsis, P. Raffai, and Z. Frei, Astrophys. J. {\bf 860}, 5 (2018).

\bibitem{vonZeipel} H. von Zeipel, Astronomische Nachrichten {\bf 183}, 345 (1910).

\bibitem{Lidov} M. L. Lidov, Planetary and Space Science {\bf 9}, 719 (1962).

\bibitem{Kozai} Y. Kozai, Astron. J. {\bf 67}, 591 (1962).

\bibitem{Smadar} S. Naoz, Annu. Rev. Astron. Astrophys. {\bf 54}, 441 (2016).

\bibitem{Miller-2002} M. C. Miller and D. P. Hamilton, Astrophys. J. {\bf 576}, 894 (2002).

\bibitem{Wen-2003} L. Wen, Astrophys. J. {\bf 598}, 419 (2003).

\bibitem{Antonini-2012} F. Antonini and H. B. Perets, Astrophys. J. {\bf 757}, 27 (2012).

\bibitem{Antonini(2017)} F. Antonini, S. Toonen, and A. S. Hamers, Astrophys. J. {\bf 841}, 77 (2017).

\bibitem{Silsbee(2017)} K. Silsbee and S. Tremaine, Astrophys. J. {\bf 836}, 39 (2017).

\bibitem{Petrovich-2017} C. Petrovich and F. Antonini, Astrophys. J. {\bf 846}, 146 (2017).

\bibitem{Liu-ApJ} B. Liu and D. Lai, Astrophys. J. {\bf 863}, 68 (2018).

\bibitem{Xianyu-2018} L. Randall and Z.-Z. Xianyu, Astrophys. J. {\bf 853}, 93 (2018).

\bibitem{Hoang-2018} B.-M. Hoang, S. Naoz, B. Kocsis, F. A. Rasio, and F. Dosopoulou, Astrophys. J. {\bf 856}, 140 (2018).

\bibitem{Liu-Quadruple} B. Liu and D. Lai, Mon. Not. R. Astron. Soc. {\bf 483}, 4060 (2019).

\bibitem{Fragione-Quadruple} G. Fragione and B. Kocsis, Mon. Not. R. Astron. Soc. {\bf 486}, 4781 (2019).

\bibitem{Fragione-nulearcluster} G. Fragione, E. Grishin, N. W. C. Leigh, H. B. Perets, and R. Perna, Mon. Not. R. Astron. Soc. {\bf 488}, 47 (2019).

\bibitem{Zevin-2019} M. Zevin, J. Samsing, C. Rodriguez, C.-J. Haster, and E. Ramirez-Ruiz, Astrophys. J. {\bf 871}, 91 (2019).

\bibitem{Liu-HierarchicalMerger} B. Liu and D. Lai, Mon. Not. R. Astron. Soc. {\bf 502}, 2049 (2021).

\bibitem{Michaely-2019} E. Michaely and H. B. Perets, Astrophys. J. Lett. {\bf 887}, L36 (2019).

\bibitem{Michaely-2020}E. Michaely and H. B. Perets, Mon. Not. R. Astron. Soc. {\bf 498}, 4924 (2020).

\bibitem{LISA} P. Amaro-Seoane, H. Audley, S. Babak, J. Baker, E. Barausse, P. Bender, E. Berti, P. Binetruy et al.,
arXiv:1702.00786 [astro-ph.IM] (2017).

\bibitem{TianQin} J. Luo, L.-S. Chen, H.-Z. Duan, Y.-G. Gong, S. Hu, J. Ji, Q. Liu, J. Mei, V. Milyukov, M. Sazhin, C.-G. Shao, V. T.
Toth, H.-B. Tu, Y. Wang, Y. Wang, H.-C. Yeh, M.-S. Zhan, Y. Zhang, V. Zharov, and Z.-B. Zhou, Classical and Quantum Gravity {\bf 33}, 035010 (2016).

\bibitem{TaiJi} W.-R. Hu and Y.-L. Wu, National Science Review {\bf 4}, 685 (2017).

\bibitem{DECIGO} T. Nakamura, M. Ando, T. Kinugawa, H. Nakano, K. Eda, S. Sato, M. Musha, T. Akutsu, T. Tanaka,
N. Seto, N. Kanda, and Y. Itoh, Progress of Theoretical and Experimental Physics {\bf 2016}, 093E01 (2016).

\bibitem{DeciHZ} M. Arca Sedda, C. P. L. Berry, K. Jani, P. AmaroSeoane, P. Auclair, J. Baird, T. Baker, E. Berti et al., arXiv:1908.11375 [gr-qc] (2019).

\bibitem{TianGo} K. A. Kuns, H. Yu, Y. Chen, and R. X. Adhikari, Phys. Rev. D {\bf 102}, 043001 (2020).

\bibitem{Tokovinin} A. Tokovinin, S. Thomas, M. Sterzik, and S. Udry, Astron. Astrophys. {\bf 450}, 681 (2006).

\bibitem{Raghavan} D. Raghavan, H. A. McAlister, T. J. Henry, D. W. Latham, G. W. Marcy, B. D. Mason, D. R. Gies, R. J. White,
and T. A. ten Brummelaar, Astrophys. J. Suppl. Ser. {\bf 190}, 1 (2010).

\bibitem{Fuhrmann} K. Fuhrmann, R. Chini, L. Kaderhandt, and Z. Chen, Astrophys. J. {\bf 836}, 139 (2017).

\bibitem{Naoz-2017} S. Naoz, G. Li, M. Zanardi, G. C. de El{\'\i}a, and R. P. Di Sisto, Astron. J. {\bf 154}, 18 (2017).

\bibitem{Chiang-2018} B. R. Vinson and E. Chiang, Mon. Not. R. Astron. Soc. {\bf 474}, 4855 (2018).

\bibitem{Naoz-2019} G.~C. de El{\'\i}a, M. Zanardi, A. Dugaro, S. Naoz, Astron. Astrophys. {\bf 627}, A17 (2019).

\bibitem{Suto-1} T. Hayashi, S. Wang, and Y. Suto, Astrophys. J. {\bf 890}, 112 (2020).

\bibitem{Suto-2} T. Hayashi and Y. Suto ,Astrophys. J. {\bf 897}, 29 (2020).

\bibitem{Suto-3} T. Hayashi and Y. Suto, Astrophys. J. {\bf 907}, 48 (2021).

\bibitem{Ransom} S. M. Ransom, I. H. Stairs, A. M. Archibald, J. W. T. Hessels, D. L. Kaplan, M. H. van Kerkwijk, J. Boyles, A. T. Deller,
S. Chatterjee, and A. Schechtman-Rook et al., Nature {\bf 505}, 520 (2014).

\bibitem{Thompson} T. A. Thompson, C. S. Kochanek, K. Z. Stanek, C. Badenes, R. S. Post, T. Jayasinghe, D. W. Latham, A. Bieryla, G. A. Esquerdo,
P. Berlind, M. L. Calkins, J. Tayar, L. Lindegren, J. A. Johnson, T. W. S. Holoien, K. Auchettl, and K. Covey, Science {\bf 366}, 637 (2019)

\bibitem{Eisner} N.~L. Eisner, C. Johnston, S. Toonen, A.~J. Frost, S. Janssens, C.~J. Lintott, S. Aigrain, et al.,
Mon. Not. R. Astron. Soc. {\bf 511}, 4710 (2022).

\bibitem{Liu-Yuan} B. Liu, D. Lai, and Y.-F. Yuan, Phys. Rev. D {\bf 92}, 124048 (2015).

\bibitem{Liu-2020-PRD} B. Liu, and D. Lai, Phys. Rev. D {\bf 102}, 023020 (2020).

\bibitem{Ford} E. B. Ford, B. Kozinsky, and F. A. Rasio, Astrophys. J. {\bf 535}, 385 (2000).

\bibitem{Naoz-PN}  S. Naoz, B. Kocsis, A. Loeb, and N. Yunes, Astrophys. J. {\bf 773}, 187 (2013).

\bibitem{Blanchet} L. Blanchet, Living Rev. Rel. {\bf 17}, 2 (2014).

\bibitem{Kiseleva-1996} L. G. Kiseleva, S. J. Aarseth, P. P. Eggleton, and R. de La Fuente Marcos, in ASP Conf. Ser. 90,
The Origins, Evolution, and Destinies of Binary Stars in Clusters, ed. E. F. Milone and J.-C. Mermilliod (San Francisco, CA: ASP), 433 (1996).

\bibitem{MD} C. D. Murray, and S. F. Dermott, Solar System Dynamics, Cambridge U. Press, NY (1999).

\bibitem{YanqingWu} Y. Wu, and P. Goldreich, Astrophys. J. \textbf{564}, 1024 (2002).

\bibitem{Lee}  M.~H.Lee, S.~J. Peale, Astrophys. J. \textbf{592}, 1201 (2003).


\bibitem{Liuetal-2015} B. Liu, D.~J. Mu{\~n}oz, and D. Lai, Mon. Not. Roy. Astron. Soc. {\bf 447}, 747 (2015).

\bibitem{Peters-1964} P. C. Peters, Phys. Rev. {\bf 136}, B1224 (1964).

\bibitem{Gaia} Gaia Collaboration, F. Arenou, C. Babusiaux, M.~A. Barstow, S. Faigler, A. Jorissen,
P. Kervella, et al., arXiv:2206.05595[astro-ph.SR] (2022).

\bibitem{Dan-Gaia} D. J. D'Orazio and A. Loeb, Phys. Rev. D, {\bf 100}, 103016 (2019).

\bibitem{Casertano-1995} S. Casertano, M. G. Lattanzi, and M. A. C. Perryman, in Future Possibilities for Astrometry in Space, ESA Special Publication,
Vol. 379, edited by M. A. C. Perryman and F. van Leeuwen (1995) pp. 47–54.

\bibitem{Bernstein} H.-H. Bernstein and U. Bastian, in Future Possibilities for Astrometry in Space, ESA Special Publication, Vol. 379, edited by
M. A. C. Perryman and F. van Leeuwen (1995) p. 55.

\bibitem{Casertano-2008} S. Casertano, M. G. Lattanzi, A. Sozzetti, A. Spagna, S. Jancart, R. Morbidelli, R. Pannunzio, D. Pourbaix, and D. Queloz,
Astron. Astrophys. {\bf 482}, 699 (2008).

\bibitem{Perryman} M. Perryman, J. Hartman, G. Á. Bakos, and L. Lindegren,  Astrophys. J. {\bf 797}, 14 (2014).

\bibitem{Ranalli} P. Ranalli, D. Hobbs, and L. Lindegren, Astron. Astrophys. {\bf 614}, A30 (2018).

\bibitem{Holmberg} J. Holmberg, C. Flynn, and L. Portinari, Mon. Not. Roy. Astron. Soc. {\bf 367}, 449 (2006).

\bibitem{GaiaFactSheet} Gaia collaboration, Expected Science Performance for the nominal and the extended mission,
\url{https://www.cosmos.esa.int/web/gaia/science-performance}.

\bibitem{Evans+2018} Gaia Collaboration, A. G. A. Brown, A. Vallenari, T. Prusti, J. H. J. de Bruijne, C. Babusiaux, C. A. L. Bailer-Jones,
M. Biermann, D. W. Evans, L. Eyer et al., Astron. Astrophys. {\bf 616}, A1 (2018).

\bibitem{Alejandro2021} A. Vigna-G\'omez, S. Toonen, E. Ramirez-Ruiz, N. W. C.
Leigh, J. Riley, and C.-J. Haster, Astrophys. J. Lett. {\bf 907}, L19 (2021).

\bibitem{XiaoFang} X. Fang, T. A. Thompson, and C. M. Hirata, Astrophys. J. {\bf 875}, 75 (2019).

\bibitem{Christian} P. Christian and A. Loeb, Mon. Not. Roy. Astron. Soc. {\bf 469}, 930 (2017).

\bibitem{Wagg} T. Wagg, F.~S. Broekgaarden, S.~E. de Mink, L.~A.~C. van Son, N. Frankel, S. Justham, arXiv:2111.13704[astro-ph.HE] (2021).

\bibitem{Kremer} K. Kremer, S. Chatterjee, K. Breivik, C. L. Rodriguez, S. L. Larson, and F. A. Rasio, Phys. Rev. Lett. {\bf 120}, 191103 (2018).

\bibitem{Xianyu-LISA} L. Randall and Z.-Z. Xianyu, arXiv:1902.08604[astroph.HE] (2019).

\bibitem{Hoang} B.-M. Hoang, S. Naoz, B. Kocsis, M. F. Will, and J. Mclver, Astrophys. J. Lett. {\bf 875}, L31 (2019).

\bibitem{Silvia-LISA} T. Robson, N. J. Cornish, N. Tamanini, and S. Toonen, Phys. Rev. D {\bf 98},064012 (2018)

\bibitem{Dan-LISA} D. J. D'Orazio and J. Samsing, Mon. Not. R. Astron. Soc. {\bf 481}, 4775 (2018).

\bibitem{Breivik}  K. Breivik, K. Kremer, M. Bueno, S. L. Larson, S. Coughlin and V. Kalogera, Astrophys. J. Lett. {\bf 854}, L1 (2018).

\end{thebibliography}
\end{document}